\documentclass[pdflatex, sn-nature]{sn-jnl}

\usepackage{graphicx}%
\usepackage{multirow}%
\usepackage{amsmath,amssymb,amsfonts}%
\usepackage{amsthm}%
\usepackage{mathrsfs}%
\usepackage[title]{appendix}%
\usepackage{xcolor}%
\usepackage{textcomp}%
\usepackage{manyfoot}%
\usepackage{booktabs}%
\usepackage{algorithm}%
\usepackage{algorithmicx}%
\usepackage{algpseudocode}%
\usepackage{listings}%

\usepackage{ragged2e}
\usepackage{array}

\theoremstyle{thmstyleone}%
\theoremstyle{thmstyletwo}%

\theoremstyle{thmstylethree}%

\begin{document}

\title[Article Title]{Dual-comb correlation spectroscopy of thermal light}


\author*[1,2]{\fnm{Eugene} \sur{Tsao}}\email{eugene.tsao@colorado.edu}

\author[1,2,3]{\fnm{Alexander} \sur{Lind}}

\author[1,2,3]{\fnm{Connor} \sur{Fredrick}}

\author[4,2]{\fnm{Ryan} \sur{Cole}}

\author[1,3]{\fnm{Peter} \sur{Chang}}

\author[2,1]{\fnm{Kristina} \sur{Chang}}

\author[2,3]{\fnm{Dahyeon} \sur{Lee}}

\author[1,3]{\fnm{Matthew} \sur{Heyrich}}

\author[2]{\fnm{Nazanin} \sur{Hoghooghi}}

\author[2]{\fnm{Franklyn} \sur{Quinlan}}

\author*[1,2,3]{\fnm{Scott} \sur{Diddams}}\email{scott.diddams@colorado.edu}

\affil[1]{\orgdiv{Department of Electrical, Computer, and Energy Engineering}, \orgname{University of Colorado, Boulder}, \orgaddress{\street{425 UCB}, \city{Boulder}, \postcode{80309}, \state{Colorado}, \country{USA}}}

\affil[2]{\orgdiv{Time and Frequency Division}, \orgname{NIST}, \orgaddress{\street{325 Broadway}, \city{Boulder}, \postcode{80305}, \state{Colorado}, \country{USA}}}

\affil[3]{\orgdiv{Department of Physics}, \orgname{University of Colorado, Boulder}, \orgaddress{\street{Libby Dr}, \city{Boulder}, \postcode{80302}, \state{Colorado}, \country{USA}}}

\affil[4]{\orgdiv{Department of Physics and Astronomy}, \orgname{Bates College}, \orgaddress{\street{2 Andrews Rd}, \city{Lewiston}, \postcode{04240}, \state{Maine}, \country{USA}}}


\abstract{The detection of light of thermal origin is the principal means by which humanity has learned about our world and the cosmos. In optical astronomy and remote sensing, direct detection of thermal photons and the resolution of their spectra have enabled discoveries of the broadest scope and impact. Such measurements, however, do not capture the phase of the thermal fields---a parameter that has proven crucial to transformative correlation techniques in radio astronomy and sensing such as synthetic aperture imaging. Over the last 25 years, tremendous progress has occurred in laser science, notably in the phase-sensitive, broad bandwidth, high resolution, and traceable spectroscopy enabled by the optical frequency comb. In this work, we directly connect the fields of frequency comb laser spectroscopy and passive optical sensing as applied to astronomy, remote sensing, and atmospheric science. We provide fundamental sensitivity analysis of dual-comb correlation spectroscopy (DCCS), whereby broadband thermal light is measured via interferometry and correlation with two optical frequency combs. We define and experimentally verify the sensitivity scaling of DCCS at black body temperatures relevant for astrophysical observations. These results allows us to provide the first evaluation of frequency comb-based thermal light spectroscopy, comparing DCCS alongside conventional laser heterodyne radiometry methods. By providing the fundamental sensitivity limits for comb-based detection and correlation of broadband thermal light, we lay a foundation for future expansion of these approaches to synthetic aperture hyperspectral imaging across the infrared and optical spectrum.}


\maketitle

For centuries, the detection of light of thermal origin across the electromagnetic (EM) spectrum has been the principle means by which humans gather information about our world and the cosmos. In astronomy, in particular, the direct detection of photons coupled with the resolution of their spectra has been key to discoveries of the broadest scope and impact \cite{ 1927ASSB...47...49L, hubble1929relation, Mayor1995, riess1998observational, perlmutter1999measurements}. However, such direct detection measurements do not take advantage of the information held in the phase of the electromagnetic field of light.

Radio astronomy instrumentation has provided a means to address this shortcoming \cite{Sullivan1982, Ryle1975}. Most significantly, the heterodyne detection and correlation of thermal EM fields relative to a common phase reference enables the reconstruction of images with improved angular resolution from arrays of telescopes arranged over long baselines \cite{Akiyama2019}.  In similar fashion, direct interferometry with light collected at distributed apertures, as originally demonstrated by Michelson \cite{Michelson1921}, yields benefits in imaging resolution, with coverage extending into the visible region of the EM spectrum.

However, these powerful phase-coherent imaging techniques have largely been disconnected from tremendous advances in the generation of, and measurement with, coherent laser light. For example, the most stable optical oscillators and clocks now have sub-cycle attosecond coherence over extended timescales \cite{Zhang2017}. Furthermore, this level of coherence and timing precision can be distributed over hundreds or even thousands of kilometers in fiber and free space \cite{Predehl2012, Shen2022, Caldwell2023}. And optical frequency combs allow one to coherently synthesize and broadcast this coherence from the radio to the optical domain, encompassing hundreds of terahertz of the EM spectrum \cite{Diddams2020}.

\begin{figure}[!ht]%
\centering
\includegraphics[width=1\textwidth]{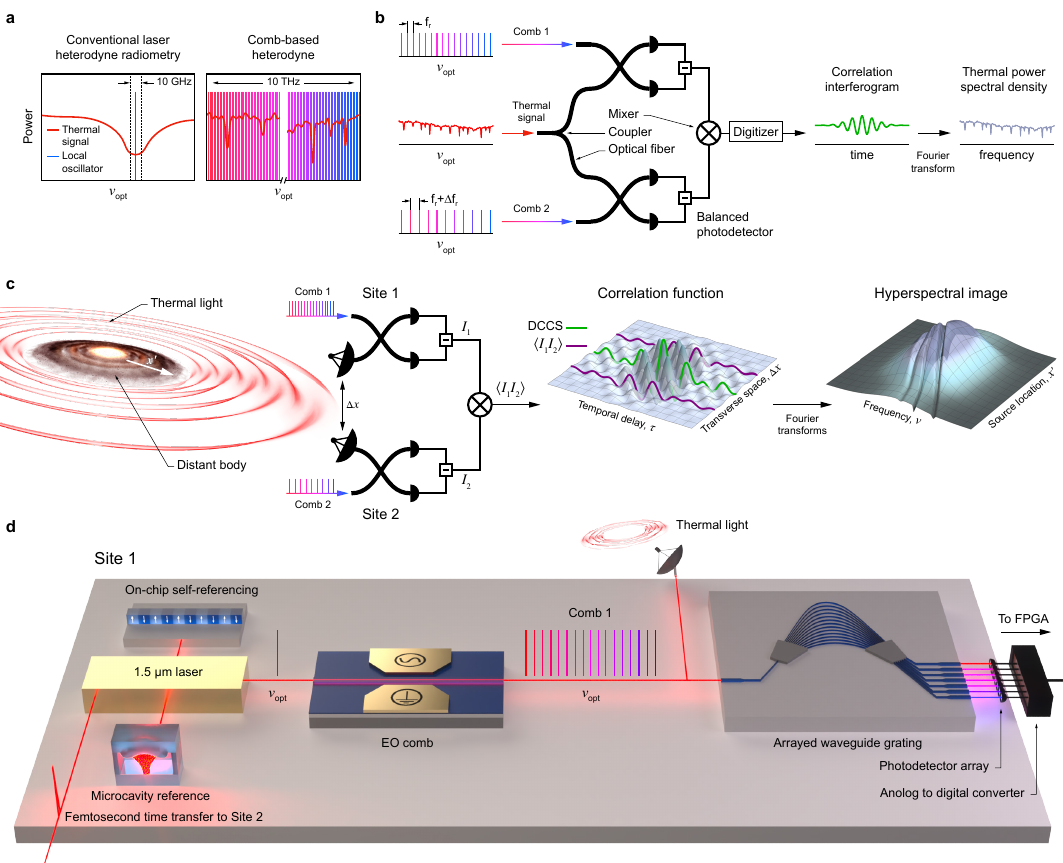}
\caption{Dual-comb correlation spectroscopy and its applications and enabling technologies. \textbf{a}, Compared to conventional laser heterodyne radiometry (LHR) with a continuous-wave local oscillator (LO), a frequency comb LO provides $1000\times$ greater bandwidth--e.g., $10+$ THz versus 10 GHz, capturing many features and photons across broad optical bandwidths \textbf{b},~In DCCS, thermal light is split then separately interfered with two mutually coherent frequency combs with offset repetition rates. The resulting photocurrents are mixed, and averaging the mixer output reveals a correlation interferogram. A Fourier transform then reveals the spectrum of the thermal light.   \textbf{c}, Dual-comb correlation is fundamentally a means to measure correlations between broadband optical fields. For example, it can measure spatial-temporal correlation functions and enable high angular resolution hyperspectral synthesis imaging. \textbf{d}, Advances in integrated photonics may enable robust and portable comb-based high-resolution spectroscopy and optical synthesis imaging. Key technologies include low SWaP-C chip-integrated combs, compact optical cavities, integrated nonlinear photonics, telecom arrayed waveguide gratings, fast photodetector arrays, heterogenous integration, high speed data processing, and femtosecond time transfer.} \label{fig:intro}
\end{figure}

In this paper, we seek to harness these revolutionary attributes of laser light and understand their optimal application to the phase-coherent heterodyne detection and correlation of broad-bandwidth EM fields of thermal origin. Specifically, we demonstrate and analyze the implementation of a frequency-comb local oscillator in the coherent detection and correlation of thermal light. Instead of a single local oscillator, the frequency comb provides thousands of local oscillators for heterodyne detection (Fig. \ref{fig:intro}a). This provides the opportunity for a significant expansion of the optical detection bandwidth, $\Delta \nu_o$, and the associated improvement in the quantum-limited signal-to-noise ratio constrained by $\sqrt{\Delta \nu_o}$ \cite{Zmuidzinas2003}. 

We illustrate the power of this approach through a technique we term dual-comb correlation spectroscopy (DCCS-Fig. \ref{fig:intro}b), capturing thermal light across optical bandwidths on the order of 100 GHz with balanced photodetection having only 50 MHz of electrical bandwidth.  This significant simplification is enabled by the spectral compression and correlation of thermal heterodyne signals with two frequency combs that have slightly different mode spacing.  We theoretically define the previously unknown sensitivity scaling and experimentally verify our model of DCCS on thermal light having power spectral density equivalent to a 5770 K black body (``Solar Blackbody limit"). Our work allows us to provide a first-of-its-kind synthesis and comparative analysis of frequency comb-based spectroscopy of thermal light, and thereby evaluate the complex trade-space of SNR, instrument complexity, and technological maturity between DCCS and more conventional laser heterodyne radiometry (LHR) methods.

Dual-comb correlation is essentially a means to measure the coherence between broadband optical thermal fields. Here we demonstrate correlation in time and, therefore, spectroscopy with a single aperture. However, our experimental results and analysis provide sensitivity limits and a technical context for optical frequency-comb-based thermal correlation, in general. By providing such understanding, we lay a crucial foundation for future work analyzing long-baseline and broad bandwidth hyperspectral synthesis imaging at frequencies in the range of $\sim$20 to 300 THz. We expect that this will most directly impact optical astronomical imaging \cite{eisenhauer2023advances}, facilitating sub-milliarcsecond angular resolution, high spectral resolution (R$>$100,000), broadband (10+ terahertz) observations of bright astronomical targets like nearby planets and stars \cite{2005Msngr.121....2H, parks2021interferometric, national2022origins} (Fig. \ref{fig:intro}c and Supplementary Note \ref{appx: synthesis imaging}). This work may also impact more general scenarios in the passive detection and measurement of broadband thermal light, including remote sensing of Earth-based and near-Earth objects, trace gas detection, and atmospheric science \cite{Brewer2019, Martan2014Radio}, harnessing the high spectral resolution and low size, weight, power, and cost (SWaP-C) offered by emerging photonics and electronics platforms \cite{Chang2022} (Fig. \ref{fig:intro}d).

\textbf{Prior Work} \label{sec1} Earlier work \cite{Giorgetta2010,Boudreau2012} first demonstrated the feasibility of dual-comb correlation spectroscopy at high thermal powers showing that combs with offset repetition rates could auto-correlate thermal light and produce a first-order correlation interferogram, $g^{(1)}$, and the corresponding power spectrum. These works found that at high thermal powers, the single-shot time domain signal-to-noise ratio (SNR) approached 1, and Boudreau and Genest examined the necessary conditions for this high power, unity SNR regime \cite{Boudreau2012}. However, the fundamental scaling of DCCS at realistically weak thermal powers remained unknown. Here we provide the fundamental scaling for dual-comb correlation spectroscopy at weak thermal powers (which also applies to the general case of correlation measurements with two frequency combs--see Supplementary Note \ref{appx: synthesis imaging}). We furthermore demonstrate spectroscopy on light over 1000x weaker than these prior works, at the astrophysically relevant ``Solar Blackbody limit.'' See Methods for a description of steps taken to improve sensitivity. Before quantifying the fundamental sensitivity of DCCS, we outline the principles of operation.

\textbf{Principles of Dual-Comb Correlation Spectroscopy}\label{sec2} In DCCS (Fig. 2a), thermal light is split on a beam splitter into two copies, then both copies are detected in balanced heterodyne detection with two comb local oscillators with offset repetition rates $f_r$ and $f_r+\Delta f_r$. The resulting photocurrents are mixed then digitized. Understood in the time domain (Fig. 2b,c), the two frequency combs sample the electric field of the thermal radiation at different delays whose interval is set by the repetition rates and their difference $\Delta f_r/f_r^2$ \cite{Coddington2016}. Mixing correlates the thermal field at these different delays, and averaging successive acquisitions (each of duration $1/\Delta f_r$) results in a first order-type correlation interferogram $g^{(1)}$. The Fourier transform of this interferogram results in the power spectral density (PSD) of the thermal light multiplied by the dual-comb spectrum \cite{Boudreau2012, Giorgetta2010}. An analogous process of spatial correlation to measure spatial coherence is required for aperture synthesis (Fig. \ref{fig:intro}b), and an extended discussion is provided in Supplementary Note \ref{appx: synthesis imaging}.

Now we introduce a new understanding of DCCS in the frequency domain, where DCCS is a down-conversion from an optical spectrum to an RF spectrum, mediated by a coherent multi-heterodyne dual-comb process (Fig. 2d). Note that the spectral compression $\Delta f_r/f_r$ is identical to that in conventional dual-comb spectroscopy \cite{Coddington2016}. Viewed in the frequency domain, thermal light continuously spans a section of optical frequencies, and comb teeth periodically span the same section separated by the repetition rate $f_r$ (Fig. 2e). Adjacent teeth from the two combs are spaced apart by multiples of $\Delta f_r$. By choosing a detection bandwidth of $B_f = f_r/2$, each comb tooth heterodynes with $f_r$-sized sections of thermal light \cite{Zmuidzinas2003}. This bandwidth is chosen in order to to capture as much optical power as possible while providing a unique mapping from optical frequency to radio frequency at a resolution of $f_r$. These heterodyne currents between the thermal light and the two comb teeth from the two combs appear as ``white noise'' from DC to $B_f$ (Fig. 2f). Importantly, these apparently random heterodyne currents are correlated at an offset frequency of $n\Delta f_r$. This correlation exists because the two combs are mutually coherent, and the thermal light at one detector is nearly identical to the thermal light at the other.

\begin{figure*}[!ht]%
\centering
\includegraphics[width=1\textwidth]{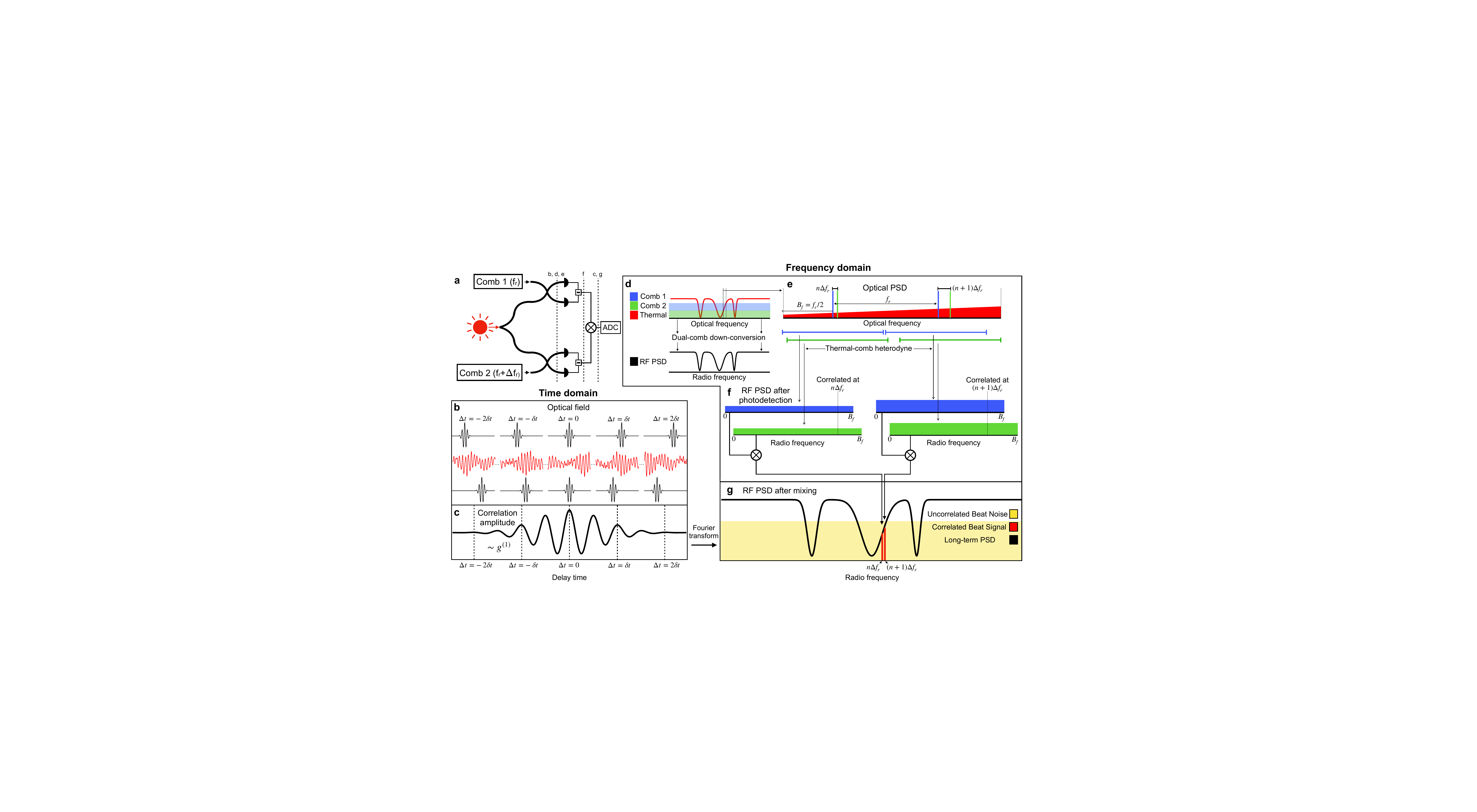}
\caption{\textbf{a},~Simplified schematic of DCCS set-up. Thermal light is split then sampled at different delays by two frequency combs. Mixing auto-correlates the thermal field. Dashed lines indicate section of experimental apparatus and corresponding sub-figure. \textbf{b},~Time domain picture of measurement. Offset repetition rates allow sampling of the thermal light at successive delays corresponding to different points on \textbf{c}, a $g^{(1)}$ type interferogram, where the real delay times can be recovered with knowledge of $f_r$ and $\Delta f_r$ as in a sampling oscilloscope or conventional DCS \cite{Coddington2016}. \textbf{d},~Frequency domain picture of DCCS. DCCS can be understood as broadband down-conversion from optical frequencies to RF frequencies. \textbf{e},~The PSD of thermal light in a given $f_r$-sized section of optical frequencies is measured by heterodyning with two comb teeth (green and blue) at offset optical frequencies. \textbf{f},~While the resulting heterodyne photocurrent for each heterodyne appears as white noise, the two heterodynes are correlated at an offset frequency set by the difference of the comb teeth from the two combs. \textbf{g},~Uncorrelated mixing results in white noise that reduces with averaging. But mixing of correlated noise produces coherent tones at offset frequencies corresponding to the difference in comb tooth frequencies (e.g. $n\Delta f_r$). The magnitudes of these tones convey the optical power of the corresponding section of spectrum.} \label{fig:Conceptual}
\end{figure*}

Next, the two heterodyne currents mix, and correlated sections of the heterodyne mix down to one offset frequency bin (Fig. \ref{fig:Conceptual}f). In this process, an $f_r$-sized section of optical frequencies is down-converted to the RF frequency $n\Delta f_r$. And the next $f_r$-sized section is down-converted to the RF frequency $(n+1)\Delta f_r$. These beats comprise the signal (represented by red blocks in Fig. 2g) and convey the optical power in each $f_r$-sized section of optical spectrum. Uncorrelated sections also mix, resulting in broadband noise represented in yellow. Averaging reduces the uncorrelated noise and maintains the signal, resulting in the power spectrum of the thermal light.

\textbf{Fundamental SNR Scaling}\label{fundamentalscaling} The statistical properties of thermal light set the maximum signal-to-noise ratio (SNR) of an estimate on the first order correlation $g^{(1)}(\tau_d) = \frac{\langle E^*(t)E(t+\tau_d)\rangle}{\langle |E(t)|^2 \rangle}$, where $E$ is the field, $\tau_d$ is the relative delay, and $t$ is time. For any kind of light, the maximum SNR of a single-shot measurement of $g^{(1)}(\tau_d)$ is the expectation value of the field autocorrelation divided by the standard deviation of the field autocorrelation. Simplification results in: $\text{SNR}_\text{max} = g^{(1)}(\tau_d)/\sqrt{g^{(2)}(\tau_d)-g^{(1)}(\tau_d)^2}$ (see Supplementary Note  \ref{secA1statisticalSNR}). The single-shot SNR is constrained by the second order correlation function, defined as $g^{(2)}(\tau_d) = \frac{\langle I^*(t)I(t+\tau_d)\rangle}{\langle I(t) \rangle^2}$, where $I \propto |E|^2$. 

For a thermal state, $g^{(2)}(0)=2$ and $g^{(1)}(0)=1$. Thus, the single-shot SNR ($\text{SNR}_\text{max}$) is 1 at the center of the correlation function. In DCCS, this value corresponds to a measurement of $V_{\text{pk}}/V_{\text{rms}}$ (interferogram envelope voltage divided by the voltage standard deviation) at the zero delay point, which is $\Delta t = 0$ in  Fig. \ref{fig:Conceptual}b and \ref{fig:Conceptual}c, over a timescale of $1/f_r$, equivalent to a resolution bandwidth of $\sim f_r$. We note for the purpose of this measurement, thermal light is considered to be chaotic light \cite{Loudon1983}, where the statistical properties that define the SNR of DCCS arise from the chaotic nature of the light and not necessarily the black-body origin (see App. \ref{thermallight}).

While a single-shot measurement of $g^{(1)}$ of thermal light is limited to 1 by the intrinsic statistics of thermal light, other sources of noise further degrade the single-shot SNR such as technical noise, and most fundamentally, photon shot noise from the comb LO. Here we provide an abbreviated derivation of the shot noise and technical noise-limited SNR. A more detailed version of the following derivation and a complete frequency domain derivation can be found in Supplementary Note \ref{ap:DCCSNRDer}.

The mean square heterodyne current power spectral density (PSD) in units A$^2$/Hz between thermal light and a frequency comb is:
\begin{equation}
\langle i_h^2 \rangle = 4  (\frac{\eta e}{h \nu})^2 S P_c,
\end{equation}
where $\eta$ is the detector quantum efficiency, $e$ is the elementary charge, $h \nu$ is the photon energy, $S$ is the PSD of the thermal light in units W/Hz, and $P_c$ is the comb power in units W. The shot noise current PSD in units A$^2$/Hz is:
\begin{equation}
\langle i_s^2 \rangle = 2 \frac{ \eta e^2 }{h \nu} P_c.
\end{equation}
The current out of the balanced detector is $i_t = i_h + i_s$. The current after mixing is then:
\begin{equation}
i_{t,1}i_{t,2}^* = i_{h,1}i_{h,2}^* + i_{s,1}i_{h,2}^* + i_{h,1}i_{s,2}^* + i_{s,1}i_{s,2}^*,
\end{equation}
where subscript 1 and 2 denote current from the two different combs. The product term of the thermal heterodyne currents is signal while all terms including the thermal heterodyne product contribute to noise. Hence, the shot noise-limited single-shot SNR is:
\begin{equation}
\text{SNR}_{\text{sn}} = \frac{\langle i_h^2 \rangle}{\langle i_h^2 \rangle + \langle i_s^2 \rangle} = \frac{\eta \langle n \rangle}{\eta \langle n \rangle + 1},
\end{equation}
where $\langle i_h^2 \rangle = \langle i_{h,1}i_{h,2}^* \rangle$ and $\langle i_h^2 \rangle + \langle i_s^2 \rangle = \langle i_{t,1}i_{t,2}^* \rangle$. $\langle n \rangle = (\exp{[h\nu/kT]}-1)^{-1}$ or the mean photon occupancy in a single polarization and spatial mode, $h\nu$ is the photon energy, $k$ is the Boltzmann constant, $T$ is temperature, and $S = h\nu \langle n \rangle$. 
While this is the shot noise limit, additional measurement noise can be considered as 
\begin{equation}
\text{SNR}_{\text{tech}} = \frac{\eta \langle n \rangle}{\eta \langle n \rangle + \chi},
\end{equation}
where $\chi$ is a factor greater than 1 describing additional noise such as electronic noise, excess relative intensity noise (RIN), or degradation of the signal strength.

\textbf{Spectroscopy at the Solar Blackbody Limit and Experimental Verification of SNR Scaling}\label{sec3} We seek to validate the achievable SNR with two frequency combs ($f_r = 100$ MHz) that are heterodyned with thermal light (App. \ref{thermallight}) generated by amplified spontaneous emission (ASE) from an unseeded optical amplifier. The overall spectral bandwidth is set by 1 nm bandpass filters, and a hydrogen cyanide (HCN) cell is placed in the path of the thermal light to impart a sharp absorption feature. Full details are provided in Methods.

\begin{figure*}[]%
\centering
\includegraphics[width=1\textwidth]{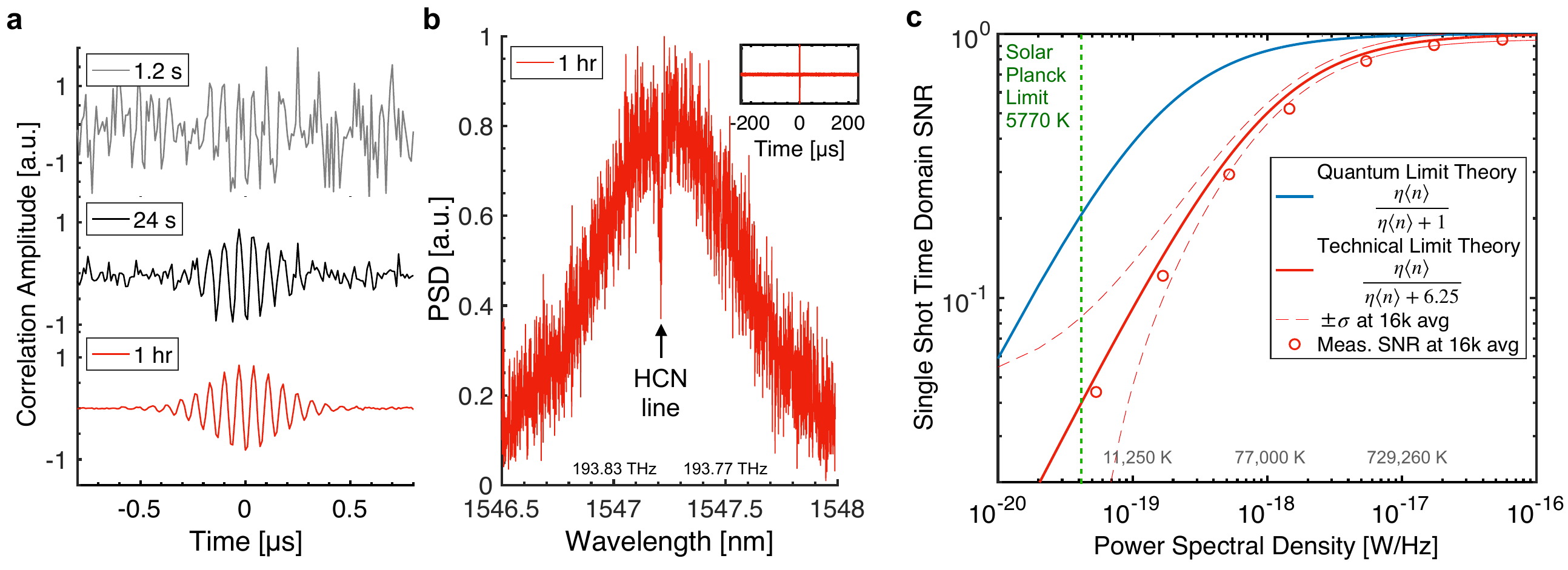}
\caption{\textbf{a}, DCCS  interferograms captured with thermal light at the PSD of $4\times 10^{-20}$ W/Hz, equivalent to a 5770 K black body at 1550 nm. Interferograms are displayed at increasing averaging times. \textbf{b}, The Fourier transform of the interferogram with 1 hr averaging yields the thermal PSD at 100 MHz resolution. The molecular absorption of a HCN line is imprinted on the thermal light. The inset shows the full interferogram window over $1/\Delta f_r \approx 500~\mu\text{s}$. The achieved spectral SNR and the theoretically estimated SNR from Eqn. 8 match at a value of 10. \textbf{c}, Theoretical and  experimental scaling of the \textit{single-shot} SNR (see Fundamental SNR Scaling). The SNR of DCCS is measured (red circles) across three orders of magnitude of PSD. The trend shows close agreement with the theoretically predicted SNR (red line) based on measurements of technical noise and detector saturation. Note, the theoretically predicted SNR is not fitted to the measured SNR. The shot noise limit (blue line) is also shown.  }\label{fig:data}
\end{figure*}

The power spectral density (PSD) of sunlight sets a benchmark for realistic sensitivity, shown as a line labeled Solar Blackbody Limit (Fig. \ref{fig:data}c), referring to the equivalent PSD of a 5770 K blackbody at 1547 nm.  We experimentally demonstrate spectroscopy at this PSD over the course of 1 hour, showing both zoomed-in interferograms and the PSD at the full 100 MHz comb resolution obtained through a fast Fourier transform (Fig. \ref{fig:data}a and \ref{fig:data}b). At 1 hour, we clearly observe the hydrogen cyanide (HCN) absorption line (Fig. \ref{fig:data}b). This is the highest-sensitivity measurement with DCCS by several orders of magnitude and the first at a PSD corresponding to a realistic astronomical source.

Our measurement exhibits a $\chi = 6.25$ that we attribute to an electronic noise floor that is $\sim$ 3 dB greater than shot noise and a heterodyne power level that is 5 dB lower than ideal. This second factor is likely due to a combination of spectral mode matching and detector saturation (see Methods). We show close agreement between the measured SNR (circles) and the theoretical technical limit (red curve) across three orders of magnitude of thermal PSD at 1547 nm (Fig. \ref{fig:data}c). Note that the theoretical technical limit does not depend on parameters fitted to the measured SNR. The experimental SNR is estimated by dividing the peak value of the averaged interferogram envelope by the root mean square value at time delays well beyond the center burst (see Methods). Hence, at lower SNR, the estimate on SNR has greater uncertainty as shown by the $\pm \sigma$ standard deviation at 16,000 averages. Clearly, reaching the shot noise (quantum) limit would yield a substantive improvement over the current technical limit (Fig. \ref{fig:data}c), and we address reaching that regime in the discussion.

\textbf{Comparing Comb-Based Thermal Light Detection}\label{sec5} Heterodyne techniques such as LHR and DCCS, while not as sensitive as direct detection, are essential for applications requiring phase sensitivity such as synthetic aperture imaging. See Supplementary Note \ref{supp: comparison} for a detailed SNR comparison of direct and heterodyne detection methods, including Echelle spectrographs, grating monochromators, and Michelson interferometer-based Fourier transform spectrometers, alongside LHR and DCCS.

Here we compare implementations of LHR and DCCS to understand the complex trade-space of sensitivity (SNR), technological maturity, and realistic technical practicalities. First, we compare the quantum-limited frequency domain SNR for these techniques, listed in Table \ref{table:SNR}, where $\Delta \nu$ is the spectral resolution in hertz, $\tau$ is the total averaging time in seconds, and $N$ is the number of resolved frequency bins per detector. Note that $N$ is a measure of optical to electronic spectral compression, which is crucial when measuring 10s of THz wide optical spectra.

\begin{figure*}[]%
\centering
\includegraphics[width=.8\textwidth]{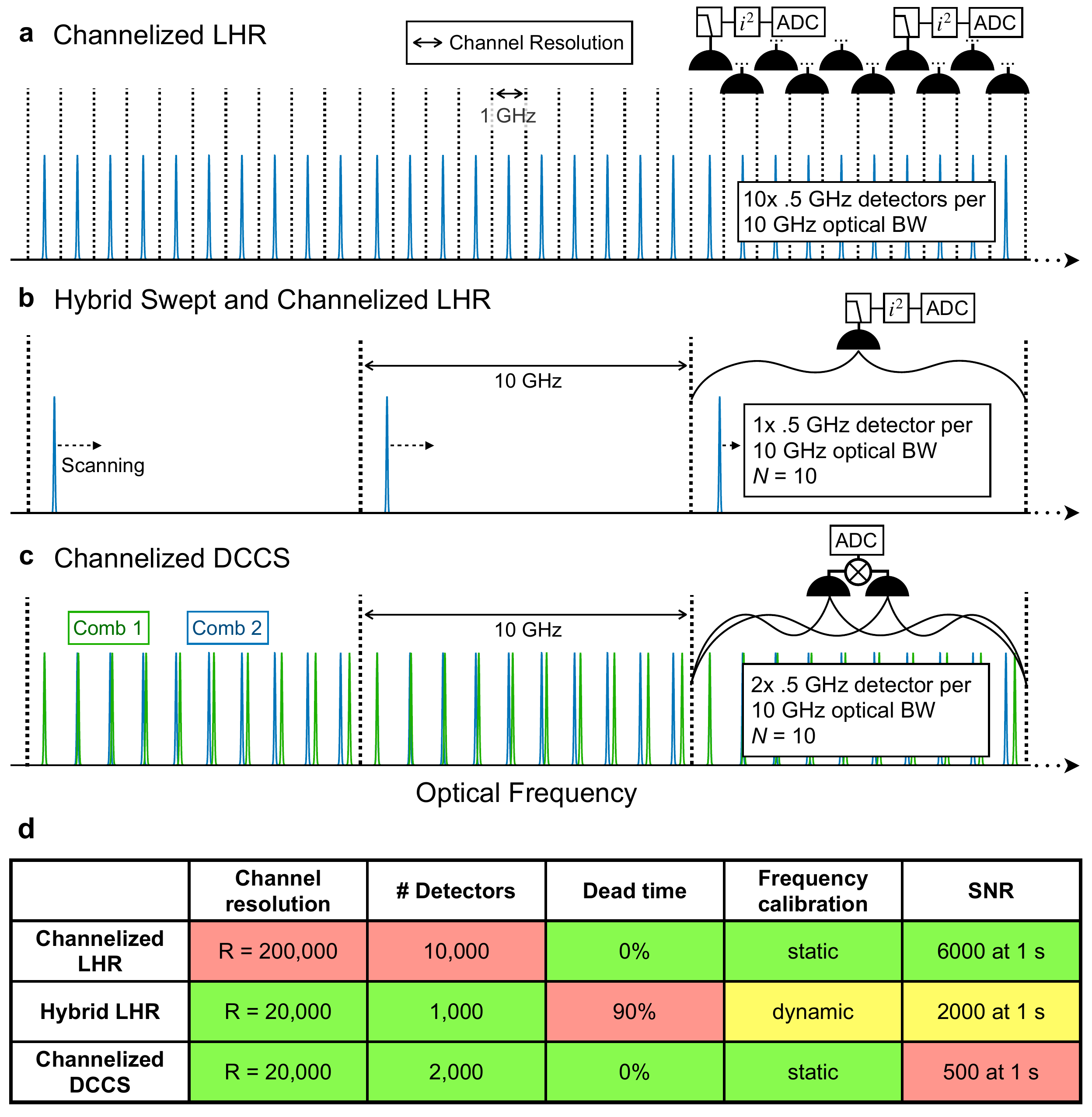}
\caption{Comparison of comb-based thermal light detection. \textbf{a}, Channelized LHR. \textbf{b}, Hybrid swept and channelized LHR. \textbf{c}, Channelized DCCS. Note that \textbf{a}, \textbf{b}, and \textbf{c} depict only 0.3\% of the total example optical bandwidth.  \textbf{d}, Comparison of techniques across channel resolution, number of detectors, dead time, frequency calibration, and SNR--see main text for parameters used. Green, yellow, and red highlighting indicates best, intermediate, and worst performance, respectively.  }\label{fig: LHR vs DCCS}
\end{figure*}

In channelized LHR (Table 1 and Fig. \ref{fig: LHR vs DCCS}a), each $\Delta \nu$-sized spectral section of thermal light is dispersed onto a single detector; thus $N$ is always 1, and the SNR does not scale with $N$ by definition (Eqn. 6). In swept LHR (Fig. \ref{fig: LHR vs DCCS}b), a single detector captures a wide band of thermal spectra as the CW local oscillator is swept in frequency, but the SNR is lowered by $1\sqrt{N}$ versus channelized LHR due to spectral dead time (Eqn. 7). DCCS (Fig. \ref{fig: LHR vs DCCS}c) captures the full broadband thermal spectrum simultaneously on a single detection channel, coming at a cost of $1/\sqrt{2}N$ over channelized LHR (Eqn. 8).

\begin{table}[]
  \centering
  \begin{tabular}{>{\centering\arraybackslash}m{6cm}>{\centering\arraybackslash}m{6cm}}
    \textbf{Spectroscopy Type} & \textbf{Frequency Domain SNR} \\
    Channelized Laser Heterodyne Radiometry & 
    \begin{equation} \frac{\eta \langle n \rangle}{\eta \langle n \rangle + 1} \sqrt{\Delta \nu \tau} \end{equation}\label{eqn:cLHR} \\
    Swept Laser Heterodyne Radiometry & \begin{equation}  \frac{\eta \langle n \rangle}{\eta \langle n \rangle + 1} \frac{1}{\sqrt{N}} \sqrt{\Delta \nu \tau} \end{equation}\label{eqn:hLHR} \\
    Dual-Comb Correlation Spectroscopy & \begin{equation} \frac{\eta \langle n \rangle}{\eta \langle n \rangle + 1} \frac{1}{N} \sqrt{\frac{\Delta \nu \tau}{2}} \end{equation}\label{eqn:cDCCS} \\
  \end{tabular}
  \caption{Quantum-limited frequency domain SNR of channelized LHR, swept LHR, and channelized DCCS. $\Delta \nu$ is the optical resolution in Hz, $\tau$ is the averaging time in s, and $N$ is the number of resolved frequency bins per detector. Other variables defined in text.}\label{table:SNR}
\end{table}

In practice, due to the aforementioned $1/\sqrt{N}$ or $1/N$ penalty of spectral multiplexing on a single detection channel, some degree of channelization is necessary for measurement across broad optical bandwidths. For a realistic comparison of implementations, we compare the following three cases: channelized LHR, hybrid swept and channelized LHR, and channelized DCCS.

\textit{Channelized LHR} (Fig. \ref{fig: LHR vs DCCS}a) is a compelling candidate for the most sensitive measurements due to its high SNR \cite{Ireland2014}. As mentioned earlier, there is no SNR penalty for increasing the number of resolved spectral bins (increasing optical bandwidth measured). However, this can only be accomplished by matching one photodetection chain to each spectral bin. For high resolution detection over broad optical bandwidths, this requirement presents significant technical challenges. For example, consider a 10 THz (80 nm at 1550 nm) section of optical bandwidth over which one desires 1 GHz resolution. A $f_r = 1$ GHz frequency comb may span this bandwidth; however, this measurement would also require a complex, low loss system for de-multiplexing 10,000 channels built on a high resolving power (R = 200,000) spectrometer. In addition, this would require 10,000 balanced detectors, rectification circuits, and digitization channels.

\textit{Hybrid Swept and Channelized LHR} is an attractive alternative both to pure channelization (above) or sweeping a single CW laser across 10 THz of optical bandwidth and introducing large amounts of dead time. As pictured in Fig. \ref{fig: LHR vs DCCS}b, a hybrid LHR approach alleviates the number of detectors and decreases the required resolving power of the spectrometer and de-multiplexing complexity as compared to channelized LHR. For example, addressing the same measurement situation as before, this approach would consist of a 10 GHz resolution spectrometer separating 10 THz total bandwidth of light onto 1000 (vs. 10,000) photodetection and digitization channels, coming at a cost of $1/\sqrt{10}$ in SNR over the channelized approach. To match this grid, one might utilize 1000 CW lasers or a $f_r$ = 10 GHz comb with 10 GHz tunability of the offset frequency. Presently the operation and active calibration of 1000 simultaneously swept CW lasers appears unrealistic. However, a frequency comb would reduce the number of lasers from 1000 to 1 and ease instrument complexity. Such tunable, high $f_r$ frequency combs are in development in the context of  astronomical spectrograph calibration \cite{Ninan2019, Sekhar2024}.

\textit{Channelized DCCS} (Fig. \ref{fig: LHR vs DCCS}c) increases instrumental simplicity compared with the LHR methods above. Channelized DCCS with 10 GHz channels measuring 10 THz of optical bandwidth with 1 GHz resolution would offer a significant simplification over pure channelized LHR--requiring, for example,  1000 instead of 10,000 digitization and detection channels. Moreover, compared to hybrid swept and channelized LHR, a channelized DCCS approach could utilize more mature frequency comb platforms with 1 GHz repetition rates, allow for static calibration (such as through a conventional dual-comb measurement \cite{Coddington2016}), and host no dead time. See Fig. \ref{fig: LHR vs DCCS}d for a comparison of channelized LHR, hybrid LHR, and channelized DCCS across channel resolution, number of detectors, dead time, frequency calibration, and SNR. Note that the SNR is calculated with parameters $\eta = 1$, $T = 5770$ K, $\lambda = 1550$ nm, $\Delta \nu = 1$ GHz, $t = 1$ s, and $N = 10$.

We highlight that in broad spectrum measurements of black-body radiation, much of the spectrum does not convey desired information, i.e., much of the spectrum consists of a smooth Planck law profile or absorption features not of interest. A spectrally tailored DCCS approach, where the combs contain power solely in spectral regions matching the absorption lines of interest, would grant even broader bandwidth measurements on a single detection channel while maintaining high SNR, which would otherwise be degraded by high spectral multiplexing, $N$, if continuously measuring broad optical spectra.

Digitization requirements for LHR are low due to the analog rectification that converts the broadband noise to a measure of DC power \cite{Fredrick2022}. We note that the digitization requirements of hybrid DCCS are also relatively easy. As a dual-comb method that maps an optical spectrum to an RF spectrum, the two combs can be tuned to map the optical spectrum to a low frequency and narrowband RF spectrum. Ultimately, this compression is limited by the mutual stability of the two combs which dictate the lowest possible $\Delta f_r$ and thus the timescale over which phase correction is necessary. Note that spectral compression is set by the ratio $\frac{\Delta f_r}{f_r}$ \cite{Coddington2016}. In practice, we have found the two combs used in the present demonstration generate phase-stable interferograms for several seconds (without requiring phase correction). With conservative parameters of 1 kHz $\Delta f_r$ (each interferogram is 1 ms in duration), 10 GHz of optical bandwidth per channel, and $f_r=$ 1 GHz, the size of the compressed RF spectrum is 10 kHz. Multiplexed across 1000 channels with 10 THz total optical bandwidth, only a modest data collection rate of 10 MB/s is required at 4 bit digitization, or 824 GB per day.

We note that an alternative heterodyne spectroscopy method would be to directly digitize the heterodyne of a demultiplexed comb and thermal radiation, then to post-process this data (e.g., through auto-correlation or FFT such as in radio spectroscopy \cite{Klein2012}) to achieve finer resolutions. A sensitivity analysis of such a technique that also addresses image-band ambiguity would be very valuable; however, the digitization requirements would likely pose significant technical challenges at broad bandwidths. For example, digitization across 10 THz would require a sampling rate of 10+ TS/s. At a bit depth of 4, this translates to a technically challenging data acquisition rate of 5+ TB/s or 86.4+ PB per day.

DCCS offers advantages in terms of instrumental simplicity--particularly when measuring across broad bandwidths--versus both channelized and hybrid LHR, but does not offer as high SNR as these methods (Fig. \ref{fig: LHR vs DCCS}d). Despite a lower SNR, DCCS may still provide the sensitivity needed for precision astronomical measurements. Consider an example Gaussian absorption line with a width of 3 GHz and depth of 0.5 imparted on light from a 5770 K black body, and a measurement efficiency of $\eta = 0.5$. A channelized DCCS scenario such as the one above, where $N = 10$ and $\Delta \nu = 1$ GHz, would reach a frequency domain SNR of 22,000 after two hours. This SNR and resolution limit the line-center measurement to an uncertainty of $8.9$ cm/s (Supplementary Note \ref{app: lineshift}), equaling the Doppler shift that the Earth imparts to sunlight. Consider the same conditions in measuring the nearest Sun-like star, $\alpha$ Centauri A, with a 10 meter telescope. Because this telescope only partially captures the available light from this star, nine days of measurement time are required to reach the SNR-limited uncertainty of $8.9$ cm/s (Supplementary Note \ref{appx: sensitivityalphacentauri}). In addition, an array of 10 meter telescopes with the measurement parameters above measuring a body like $\alpha$ Centauri A would reach sufficient SNR to correct for atmospheric fluctuations at the 10 ms timescale (Supplementary Note \ref{appx: fringealphacentauri}).

\textbf{Discussion and Conclusion}\label{sec5} As mentioned in the experimental results, simultaneous detector noise and detector saturation presently prevent quantum-limited measurement, and in general, shot noise-limited detection of short optical pulses remains a challenge for frequency comb measurements. However, the use of highly chirped pulses, e.g. from electro-optic combs, or unamplified detectors \cite{Guay2023} could alleviate saturation issues. This would allow for increased comb power, raising shot noise above other noise sources while not attenuating the heterodyne signal.

Our experimental work operated near 1550 nm due to the availability of inexpensive, high-quality, telecommunications-compatible components such as high quantum efficiency balanced photodetectors and low-loss fiber optics. However, many other regions of the optical spectrum are of great interest, particularly further into the infrared. At longer wavelengths, mean photon occupancy $\langle n \rangle$ increases, granting the ability to measure cooler objects. However, shot noise-limited DCCS is challenging in the mid- and far-IR due to a lack of high-power, low-excess RIN combs and a lack of high-speed, low-noise, high-quantum-efficiency balanced detectors. A compelling alternative is through electro-optic sampling \cite{Kowligy2019,Benea-Chelmus2019,Riek2017}, where MIR/FIR light is up-converted to the NIR for higher efficiency, low noise detection. Further analysis of the nonlinear conversion efficiency is required to fairly assess the merits of nonlinear DCCS.

While we derived the shot noise-limited SNR for DCCS, other techniques may increase performance beyond this level. For example, recent work in compressive sensing dual-comb spectroscopy \cite{giorgetta2024free} with a time programmable frequency comb enables faster acquisition times and comparable spectral characterization, and similar methods could be employed in DCCS. The use of non-classical light \cite{Tsang2011, Gottesman2012, Brown2023} to surpass the shot noise-limited SNR is also intriguing; of particular future interest is two-mode squeezing between the two frequency combs to correlate their noise, or the injection of squeezed vacuum at the unused port of the beamsplitter used to split the thermal light.

We also note that another related technique, frequency comb ptychoscopy \cite{Benirschke2021}, has similar advantages of heterodyne-based methods for thermal light spectroscopy. An understanding of the fundamental SNR limits of frequency comb ptychoscopy, as well as data collection and processing requirements, would be valuable in assessing this method. 

In summary, we provide the foundational understanding of DCCS, a heterodyne-based spectroscopy technique that leverages the phase coherence and broadband nature of the optical frequency comb to the measurement of thermal spectra. We define the fundamental quantum limits of DCCS, verify its sensitivity scaling with thermal power, and measure spectra at the Solar Blackbody Limit. We then provide analysis of comb-based thermal light spectroscopy in general, exploring the complex trade-space of SNR, instrument complexity, and technological maturity between DCCS and conventional LHR methods. Our work not only reveals fundamental sensitivity limits of DCCS, it also provides the fundamental sensitivity of dual-comb correlation in general, thus establishing the groundwork for explorations of high-angular- and spectral-resolution broadband synthesis imaging.

\section*{Methods}\label{methods}

\begin{figure*}[]%
\centering
\includegraphics[width=1\textwidth]{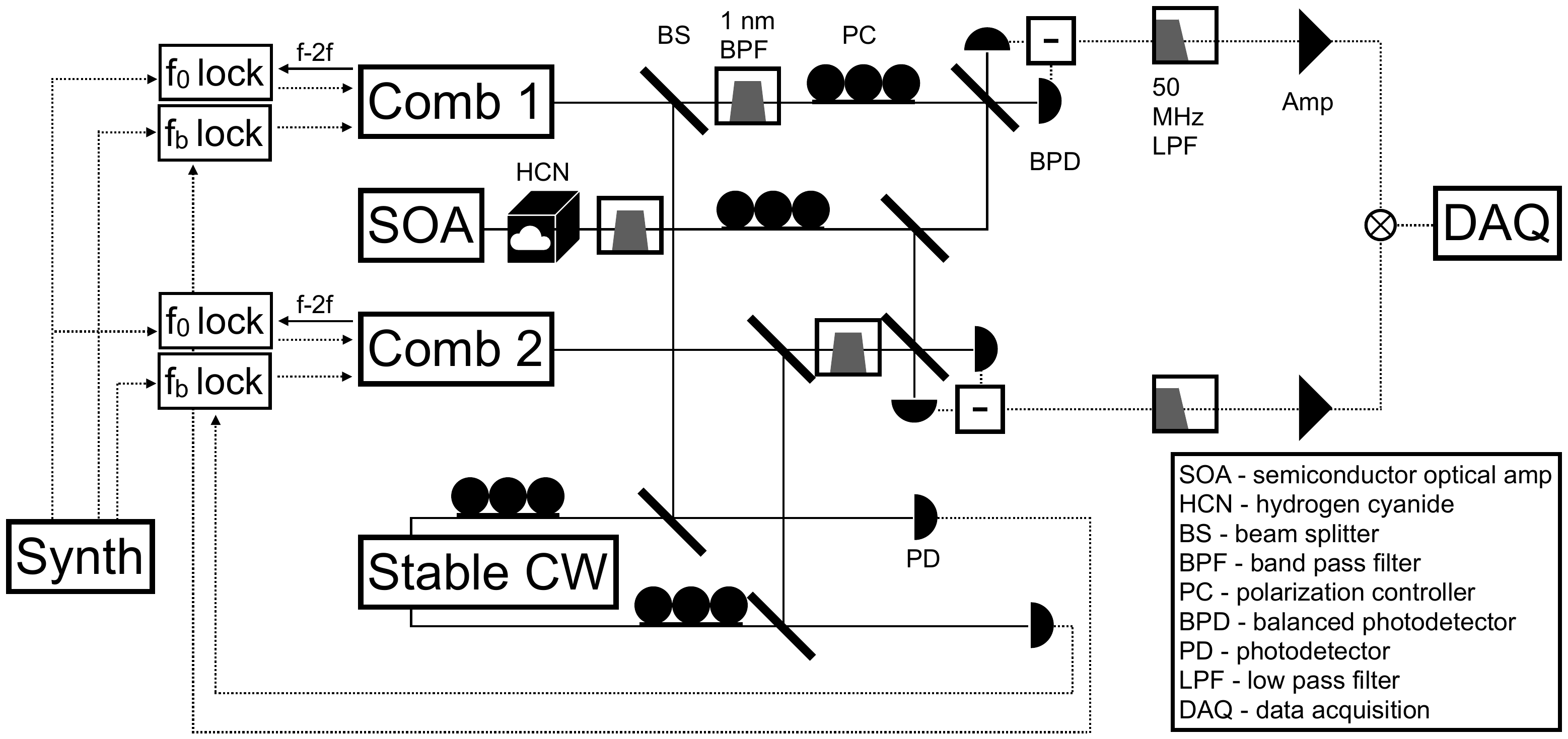}
\caption{Experimental Diagram}\label{fig: exptdiagram}
\end{figure*}

In our experiment (Fig. \ref{fig: exptdiagram}), quasi-thermal light from amplified spontaneous emission (ASE) of a semiconductor optical amplifier (SOA) is passed through a hydrogen cyanide (HCN) gas cell (NIST Standard Reference Material 2519a) to impart sharp absorption features, then is passed through a tunable 1 nm bandpass filter around 1547 nm, and is then attenuated to reach the desired PSD. After polarization control, this quasi-thermal ASE light is split and combined on a 50:50 splitter with two mutually coherent self-referenced frequency combs (Menlo Systems ULN Combs), both filtered at 1547 nm. The combined light is heterodyned via balanced detection (Thorlabs PDB410C). Comb powers approach 60 $\mu$W per detector. The outputs of the balanced detectors are low-pass filtered at 50 MHz then amplified. The two parallel heterodyne signals are mixed and the resulting current is digitized. This signal is averaged over many interferograms and then fast Fourier transformed to yield the product of the spectra of the dual-comb and thermal light.

The two combs operate at repetition rates ($f_{r,1}$, $f_{r,2}$) of approximately 100 MHz and 100 MHz + $2.083\bar{3}$ kHz ($\Delta f_{r} = 2.083\bar{3}~\text{kHz}$), set within the precision of our locking electronics. These settings result in an integer number of points per interferogram, which eases the post-processing of successive interferograms. The two combs are locked to a stable CW laser at approximately 1550 nm at the same beat frequency. Phase correction of both the interferogram carrier envelope offset and the envelope is utilized every few seconds to ensure coherent averaging. Phase correction on this time scale is required due to thermal drifts, as well as slight deviations from the exact matching of the sampling frequency, repetition rates, and $\Delta f_{r}$.

We note that the sensitivity ultimately achieved was granted through a systematic evaluation of all possible sources of loss and noise. This includes spectral and polarization mode-matching, detector saturation and nonlinearity, detector and instrument noise, digitization noise, mixer loss, mutual coherence between frequency combs, and synchronization of laser repetition rates with data acquisition rates.

The single-shot time domain SNR (Fig. 2) is evaluated by first taking the absolute value of the analytic representation of an averaged interferogram and measuring the envelope peak value. This number is then divided by the root mean square value far outside the interferogram peak to estimate the SNR. The SNR is then projected to a single shot value by dividing by the root number of interferograms. The frequency domain SNR is also evaluated in a similar fashion, since in the present measurement only a small section of the available optical Nyquist zone is used. We measure the full-width half maximum amplitude of the measured spectrum in the Fourier domain (which occupies only a few MHz) and divide this number by the level of noise outside this spectral region.

The noise and signal levels contributing to $\chi$ are measured with an Agilent MXA Signal Analyzer N9020A, with an estimated absolute uncertainty of $\pm 0.5$ dBm/Hz.

Product names are given for scientific clarity only and do not represent an endorsement by NIST.

\backmatter

\bmhead{Acknowledgements}

The authors wish to acknowledge the helpful conversations with and comments from Fabrizio Giorgetta, Igor Kudelin, John Monnier, John Lehman, Josh Combes, Noah Lordi, Jun Ye, Laura Sinclair, Jordan Wind, Jérôme Genest, Molly Kate Kreider, Will Schenken, Pooja Sekhar and Allison Duh.

\section*{Declarations}
\begin{itemize}

\item Funding: This project was supported by the W. M. Keck Foundation, the NSF through the QLCI Award No. OMA-2016244, ONR Award No. N00014-21-1-2606 and NIST. E.J.T. acknowledges support from the Office of Naval Research through the National Defense Science and Engineering Graduate Fellowship. M.J.H. acknowledges support from the NSF Graduate Research Fellowship Program.

\item Author contributions:
 
Conceptualization: E.J.T., A.J.L., S.A.D.

\noindent Methodology: E.J.T., A.J.L., C.F., F.Q., S.A.D.

\noindent Investigation: E.J.T., A.J.L., C.F., P.C., K.F.C., D.L., M. H., N. H., F.Q. S.A.D.

\noindent Visualization: E.J.T., A.J.L., R.K.C., S.A.D.

\noindent Funding acquisition: S.A.D.

\noindent Project administration: E.J.T., S.A.D.

\noindent Supervision: F.Q., S.A.D.

\noindent Writing-original draft: E.J.T., S.A.D.

\noindent Writing-review and editing: E.J.T., A.J.L., C.F., R.K.C., P.C., K.F.C., D.L., M. H., N. H., F.Q. S.A.D.

\item Conflict of interest: Authors declare that they have no conflicts of interest.

\item Data availability: Available upon request

\item Code availability: Available upon request

\item Ethics approval and consent to participate: Not applicable
\item Consent for publication: Not applicable
\item Materials availability: Not applicable

\end{itemize}

\begin{appendices}

\section{Broadband Optical Synthesis Imaging with Dual-Comb Correlation} \label{appx: synthesis imaging}

Dual-comb correlation spectroscopy measures the coherence of thermal light as a function of the temporal delay across the same spatial mode. This results in a $g^{(1)}$-type measurement of thermal light, and a Fourier transform yields the spectrum of this light. The thermal field can also be sampled and correlated at different spatial ``delays,'' then Fourier transformed to produce an image of the source. Such techniques are used for synthesis imaging in radio astronomy \cite{Thompson2017, 1999ASPC..180.....T}, and hyperspectral images are formed as image ``cubes,'' which are generated through Fourier transforms of complex visibility (coherence) as a function of the two spatial delays and the temporal delay.

Most generally, dual-comb correlation consists of the broadband phase-sensitive measurement of thermal fields with two combs followed by correlation of these signals. It is a means to measure the similarity of broadband optical fields at different points in time and space. This similarity is described as coherence or as complex visibility, and the combination of many pairwise comparisons fills out the complex visibility, which can be Fourier transformed in time and space to generate a hyperspectral image of a distant source. Here, we provide a brief description of how dual-comb correlation synthesis imaging might be implemented and how the sensitivity bounds provided in this work provide the fundamental limits for such a system.

First, we review the Fourier relationships between the complex visibility $V(u,\tau_d) = \langle E(\upsilon, t) E(\upsilon+u, t+\tau_d) \rangle$ and a hyperspectral image (modified intensity) $I(l, \omega)$. For simplicity, only one spatial (transverse) dimension ($u$ and $l$) and a temporal (longitudinal) dimension ($\tau_d$ and $\omega$) are specified. The complex visibility can first be Fourier transformed to reveal the spectrum at each spatial frequency $u$:
\begin{equation}
V(u, \omega) = \int V(u, \tau) e^{2 \pi i \tau_d \omega} d \tau,
\end{equation}
and the hyperspectral image can be recovered by taking the visibility at each spatial frequency and performing another Fourier transform:
\begin{equation}
I(l, \omega) = \int V(u, \omega) e^{2 \pi i u l} d u,
\end{equation}
where $u = \frac{\omega}{2 \pi c} \Delta x$ is the distance between sites in terms of the number of wavelengths, and $l$ is the direction cosine from a point on the source plane. These transform pairs are described by the Wiener-Khinchin theorem and the van Cittert-Zernike theorem.

\begin{figure} []
    \centering
    \includegraphics[width=1\linewidth]{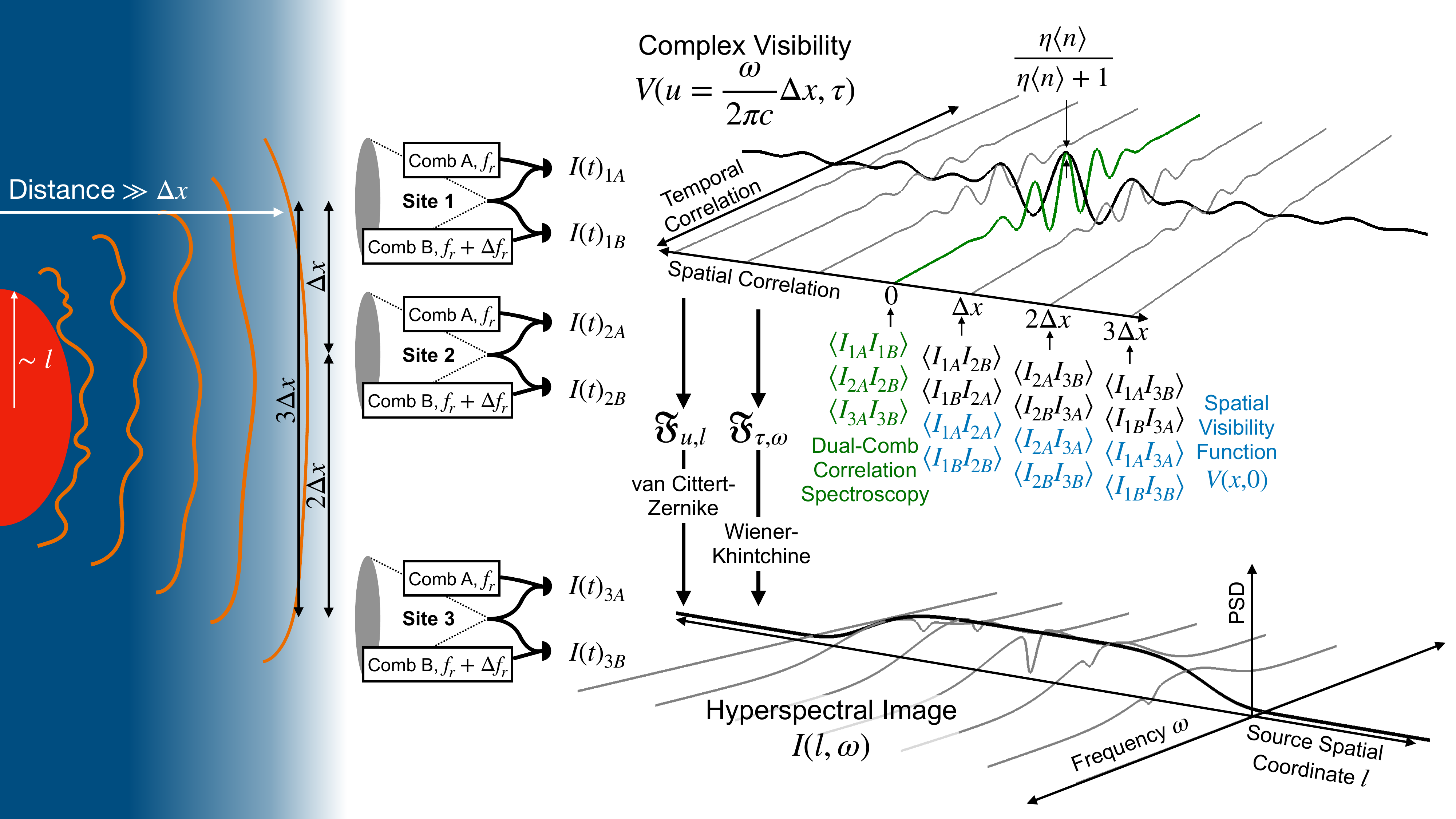}
    \caption[Optical phased array hyperspectral imaging: reconstructing the complex visibility function with dual-comb correlation]{Reconstructing the complex (spatial-temporal) visibility function with dual-comb correlation. Here a minimal three sites are represented to correlate the thermal field across three locations. Each pairwise correlation of photo-currents fills in a plane of the complex visibility. The double Fourier transform over frequency and space results in a hyperspectral image of the source.}
    \label{fig:imagesynthesis}
\end{figure}

Dual-comb correlation can sample this complex visibility (and thus generate a high-resolution hyperspectral image), and here we describe one possible implementation. The standard setup for dual-comb correlation spectroscopy is arranged at many sites, with Figure \ref{fig:imagesynthesis} showing three sites. The photocurrents between the thermal light and the frequency combs are recorded, and timing synchronization is maintained by optical time transfer, for example \cite{Caldwell2023}. These photocurrents are then pairwise correlated with others, noted as $\langle I_{x}I_{y}\rangle$, resulting in a plane-by-plane recording of the complex visibility. Such a complex visibility can be transformed as described above to form a hyperspectral image.

We note that dual-comb correlation spectroscopy records the complex visibility in the zero transverse spatial delay plane. The maximum single shot signal-to-noise ratio for dual-comb correlation of $\frac{\langle n \rangle}{\langle n \rangle + 1}$ derived and verified in this work in fact sets the upper limit for sampling the complex visibility in general. This can be seen in Fig. \ref{fig:imagesynthesis} as the zero delay point in time and space of the complex visibility. Our work shows how this sensitivity translates to the frequency domain for different thermal light powers given our spectral resolution; but this limit also bounds the spatial domain sensitivity limit. While this represents the fundamental limit, we recognize that other specific parameters of the measurement apparatus (aperture size, filling factor, etc.) and source (brightness, distance, wavelength, etc.) \cite{1999ASPC..180.....T} act to reduce the fundamental SNR (see Appendix \ref{appx: sensitivityalphacentauri}).

As with radio astronomy arrays, image synthesis of broadband light could provide much higher angular resolution than is available with today's large telescopes and Michelson stellar interferometers. The cost of large-scale telescopes is estimated to scale with a power greater than the square of the diameter \cite{VanBelle2004} quickly reaching infeasible levels; and prohibitive technical challenges occur when scaling up Michelson stellar interferometers to longer distances and especially between many sites. For example, direct detection incurs a $\approx1/\sqrt{N_t}$ sensitivity penalty due to physical splitting of light and requires a vast infrastructure of optically stable and large diameter ($>1$ meter) vacuum light pipes beyond 1 km baselines. The finer spatial resolution desired (and thus baseline length required) the more area one must “fill”, thus increasing both the telescope number $N_t$ and the light pipe scale and complexity \cite{Monnier2003}.

In contrast, a heterodyne interferometer built on emerging optical timekeeping infrastructure could scale favorably beyond km-scale arrays ($<<1$ mas), as each additional telescope site does not drastically increase total array complexity, requiring a stable oscillator (such as a laser frequency comb and high-Q optical cavity) and a fiber or free-space optical path to a neighbor. An array of smaller optical heterodyne telescopes (which can more fully cover the [u-v] plane than direct interferometry) connected and calibrated through optical time transfer may circumvent the need for large, optically aligned, monolithic telescopes and provide better cost scaling, granting a larger area of coverage to collect more photons at feasible costs. Fringe capture, however, must still be considered carefully--see Appendix \ref{appx: fringealphacentauri} for an analysis of fringe capture rates and the ability to correct for atmospheric variability between sites.

Such a system built on a dual-comb local oscillator scheme, like that shown in this paper, could be well-matched for broadband and ultra-high spatial and spectral resolution imaging of bright targets. For example, while there are extensive studies of the highly complex and dynamic behavior of our Sun that fully resolve and spectrally measure spots, convection cells, and even finer behavior, no equivalent or unambiguous observations exist for even nearby stars due to insufficient resolution and [u-v] plane coverage \cite{parks2021interferometric}. Such observations would shed light on the critical question of our solar system’s uniqueness. Furthermore, imaging bright objects within our solar system at high spatial and spectral resolutions would offer unprecedented investigations of, for example, open questions in ring, atmosphere, and core compositions and dynamics \cite{national2022origins}. A resolution of 100 µas would enable, for example, $\approx 20$-meter resolution imaging of the (bright, negative magnitude) surface of Mars, and hundreds of meter resolution imaging of Saturn and Jupiter, granting space probe-like detail but with flexible and long-term fields of view.

We also emphasize that a fully filled, broadband heterodyne array may unlock remarkable new passive‑sensing capabilities for a wide range of atmospheric, Earth‑based, and near‑Earth targets. For instance, a 10 km ground‑based array could resolve 100 µm features in the near‑infrared on low‑Earth‑orbit satellites (800 km altitude), enabling remote detection of material strain. Conversely, a 10 km satellite‑based array at the same altitude, looking down on Earth, could resolve 100 µm details in critical infrastructure, biological matter, and chemical emissions. When paired with the broad bandwidth, high resolution, and precise spectroscopy enabled by the optical frequency comb, these capabilities point toward the possibility of remote‑sensing hyperspectral meso- and perhaps microscopy.

\section{Statistical Limit of Single-Shot Time Domain Signal-to-Noise Ratio}\label{secA1statisticalSNR}

The time domain signal is the mean of the auto-correlation function of the electric field:
\begin{equation}
    A(\tau_d) = \frac{1}{t_{\text{tot}}} \int_t dt E^*(t)E(t-\tau_d),
\end{equation}
where $t_\text{tot}$ is the total time over which the measurement occurs. The time domain noise is the variance of the auto-correlation function of the electric field:
\begin{equation}
    N(\tau_d) = \frac{1}{t_{\text{tot}}} \int_t dt [E^*(t)E(t-\tau_d) - A(\tau_d)]^2.
\end{equation}
Note that $\tau_d$ is the temporal delay between sections of the field, and here we represent the expectation value $\langle x \rangle$ as the integral $\frac{1}{t_{\text{tot}}} \int_t x dt$ assuming the ergodic hypothesis, which is satisfied assuming measurement times much longer than the pulse-to-pulse interval of optical frequency combs (e.g., ns-scale). Ultimately, we care about the power spectral density determined by the time-domain interferogram, which can be expressed through the Wiener-Khinchin theorem:
\begin{equation}
   S(f) = \int_{-\infty}^{\infty} d\tau_d e^{-2 \pi i f \tau_d} R(\tau_d),
\end{equation}
where
\begin{equation}
   R(\tau_d) = A(\tau_d) \pm \sqrt{\frac{N(\tau_d)}{\mathcal{N}}}.
\end{equation}
$R(\tau_d)$ is the captured interferogram, $\sqrt{N(\tau_d)}$ is the standard deviation, and $\mathcal{N}$ is the number of trials. However, here we simply compute the time domain SNR. The frequency domain SNR is computed in a following section. As $\mathcal{N} \to \infty$, $R(\tau_d) \to A(\tau_d)$. We now analyze the case of SNR for one trial. The single shot SNR is:
\begin{equation}
   \text{SNR}_s(\tau_d) =\frac{A(\tau_d)}{\sqrt{N(\tau_d)}}
\end{equation}
Note that:
\begin{equation}
    A(\tau_d) = g^{(1)}(\tau_d) \times \langle |E(t)|^2 \rangle = g^{(1)}(\tau_d) \times \langle I \rangle.
\end{equation}
Here we define $g^{(1)}(\tau_d) = \frac{\langle E^*(t)E(t+\tau_d)\rangle}{\langle |E(t)|^2 \rangle}$, $g^{(2)}(\tau_d) = \frac{\langle I^*(t)I(t+\tau_d)\rangle}{\langle |I(t)| \rangle^2}$, and $I \propto |E|^2$. The variance is the expected value of the squared deviation from the mean:
\begin{multline}
    N(\tau_d) = \langle (E^*(t)E(t-\tau_d) -  A(\tau_d))^2 \rangle 
    \\ = \langle (E^*(t)E(t-\tau_d))^2 \rangle -  \langle 2 (E^*(t)E(t-\tau_d) A(\tau_d) \rangle + \langle A(\tau_d)^2 \rangle,
\end{multline}
simplifying:
\begin{equation}
    N(\tau_d) = \langle (I^*(t)I(t-\tau_d) \rangle - \langle A(\tau_d))^2 \rangle = \langle I \rangle^2 (g^{(2)}(\tau_d) - g^{(1)}(\tau_d)^2).
\end{equation}
Thus,
\begin{equation}
   \text{SNR}_s(\tau_d) = \frac{g^{(1)}(\tau_d)}{\sqrt{g^{(2)}(\tau_d) - g^{(1)}(\tau_d)^2}}
\end{equation}
For thermal light, we know that:
\begin{equation}
    g^{(2)}(\tau_d) = 1 + g^{(1)}(\tau_d)^2
\end{equation}
Thus,
\begin{equation}
   \text{SNR}_{s,th}(\tau_d) = \frac{g^{(1)}(\tau_d)}{\sqrt{1 + g^{(1)}(\tau_d)^2 - g^{(1)}(\tau_d)^2}} = g^{(1)}(\tau_d).
\end{equation}
At the zero delay point ($\tau_d = 0$), the single-shot SNR is 1 and falls off to 0 as a function of $g^{(1)}$ or the spectral lineshape and width.

\subsection{Thermal Light}\label{thermallight}

In this work, we follow the conventions of Mandel and Wolf for defining thermal light \cite{mandel1995optical}. Specifically, we consider thermal light any light that has a ``thermal'' distribution in phase space or Bose-Einstein distribution in intensity, which ultimately gives rise to the second-order correlation function $g^{(2)}(0) = 2$, which limits the achievable single shot SNR for dual-comb correlation to 1. These thermal statistics are a consequence of the addition of many uncorrelated fields from many uncorrelated emitters, which is the case in both ASE and emission from a black body. A detailed theoretical analysis can be found in Mandel and Wolf’s Optical Coherence and Quantum Optics, Chapter 13--Radiation from thermal equilibrium sources, Section 13.2--Thermal light \cite{mandel1995optical}. Thus, the relevant quantum statistics are the same for thermal light generated by amplified spontaneous emission vs. thermal light from a black body. We note that the spectral envelopes are, of course, different between ASE and black body radiation, however, in the case of our experiment, we filter and attenuate this light to match the intensity of a single spatial mode of light from a 5770 K black body at 1550 nm.

The statistics of ASE-derived thermal light have also been experimentally studied. Notably, Voss \textit{et al.} found that a single mode of ASE across a wide range of pumping strengths clearly displays Bose-Einstein statistics in the intensity probability distribution \cite{voss1999photon}. In their experiment, they employed homodyne tomography to reach high SNR and to select a single mode of the ASE field, which is very similar in method to DCCS, where a strong and coherent local oscillator is used.

\section{Derivation of Dual-Comb Correlation SNR vs. PSD}\label{ap:DCCSNRDer}

In order to derive the SNR of DCC on thermal light we first find the heterodyne powers between a coherent LO (CW or comb) and thermal light. Note that we interchangeably use the terms ASE, thermal light, and chaotic light.

\paragraph{CW laser and Thermal Light Heterodyne}\label{CWhet}

Before deriving the heterodyne power between a comb LO and thermal light, we first re-derive the heterodyne power between a CW LO and thermal light. To our knowledge, this equation was first derived by N.A. Olsson in 1989 \cite{Olsson1989} and A. Yariv in 1990 \cite{Yariv1990}. Both authors were motivated by the presence of amplified spontaneous emission in optical communications.

\begin{figure}[]
\centering
\includegraphics[width=.5\textwidth]{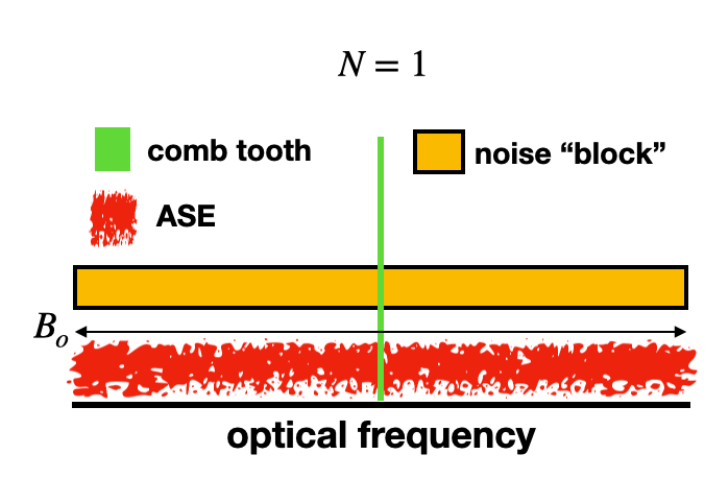}
\caption{Frequency domain picture of ASE and single comb tooth (CW).}\label{fig: supplementdiagram}
\end{figure}

Following Olsson, we first rederive the heterodyne level of a CW laser with ASE in our own notation. We will maintain this notation as we derive the comb and ASE case as well as the dual-comb correlation measurement of ASE. $P_{t}$ is the power of the CW laser or a single comb tooth. $S$ is the power spectral density of the ASE or quasi-thermal light source. $B_o$ is the total optical bandwidth over which the comb tooth heterodynes with the ASE light, equivalent to twice the detector bandwidth $B_f$ (see Fig. \ref{fig: supplementdiagram}).

We represent the electric field (expressed in root power) of the ASE light as:
\begin{equation}
    E_{\text{ASE}}(t) = \sum_{k = -B_o/2\delta \nu}^{B_o/2\delta \nu} \sqrt{2 S \delta \nu} \times \cos((\omega_o + 2 \pi k \delta \nu)t + \Phi_k),
\end{equation}
where the average power in a chunk of ASE is $S \delta \nu$, and $\delta \nu$ is a small frequency slice. $\omega_o$ is the optical frequency of one comb tooth. $\Phi_k$ is a random phase. $t$ is time. Note that the mean square of the electric field of the ASE is the average power: $S B_o = P_{\text{avg, ASE}}$. The electric field of the CW laser or one tooth of the comb light is:
\begin{equation}
    E_t(t) = \sqrt{2 P_t} \cos(\omega_o t + \Theta).
\end{equation}
$\Theta$ is the phase of this comb tooth. Note that the mean square of the comb tooth/CW field is also the average power: $P_t = P_{\text{avg, CW}}$.
The optical power incident on the detector is then the square of the sum of the two fields:
\begin{equation}
    P(t) = (\sum_{k = -B_o/2\delta \nu}^{B_o/2\delta \nu} \sqrt{2 S \delta \nu} \times \cos((\omega_o + 2 \pi k \delta \nu)t + \Phi_k) + \sqrt{2 P_t} \cos(\omega_o t + \Theta))^2.
\end{equation}

We ignore the products between the laser with itself and the ASE with itself. Note the laser self-heterodyne will result in a DC component and a very high optical frequency component, both of which can be ignored. We will assume a large LO limit for the laser, so the self heterodyne of the ASE will be negligible compared to the laser-ASE heterodyne. We then ignore the sum components of the heterodyne oscillating at frequencies around $2 \omega$ far beyond our detector bandwidth. The resulting optical power for the difference components is:
\begin{equation}
    2 \sqrt{2S \delta \nu 2 P_t} \sum_{k = -B_o/2\delta \nu}^{B_o/2\delta \nu} \frac{1}{2} \cos(2 \pi k \delta \nu t + \Phi_k - \Theta).
\end{equation}
The photocurrent is then:
\begin{equation}
    i(t) = \frac{\eta e}{h \nu}\sqrt{2S \delta \nu 2 P_t} \sum_{k = -B_o/2\delta \nu}^{B_o/2\delta \nu} \cos(2 \pi k \delta \nu t + \Phi_k - \Theta),
\end{equation}
where $\eta$ is the quantum efficiency of the detector, $e$ is the elementary charge, and $h \nu$ is the energy per photon. We're interested in the noise power of the photocurrent, and when multiplying the cosine terms of the sum, we add each product in quadrature due to the random phases of $\Phi_k$.
\begin{equation}
    \langle i(t)^2 \rangle =  \biggl( \frac{\eta e}{h \nu}\sqrt{2S \delta \nu 2 P_t} \biggr)^2 \times \Biggl(\sqrt{2\times \frac{B_o}{2\delta \nu}}\Biggr)^2 \times \frac{1}{2}.
\end{equation}
The second product term is the square root (due to quadrature addition) of the number of products between the different cosine functions. The third product term, $\frac{1}{2}$, is the average of the square of cosine. We simplify further:
\begin{equation}
    \langle i(t)^2 \rangle = \frac{\eta^2 e^2}{(h \nu)^2} \times 4S \delta \nu P_t \times 2\times \frac{B_o}{2\delta \nu} \times \frac{1}{2}.
\end{equation}
We then cancel terms terms:
\begin{equation}
    \langle i(t)^2 \rangle = 2\frac{\eta^2 e^2}{(h \nu)^2} S P_t \times B_o.
\end{equation}
This current folds uniformly into a bandwidth of $B_o/2$, hence we divide by this bandwidth to extract the power spectral density:
\begin{equation}
    \langle i(t)^2 \rangle_{\text{PSD,CW}} = 4\frac{\eta^2 e^2}{(h \nu)^2} S P_t.
\end{equation}

\paragraph{Comb and Chaotic Light Heterodyne}

For the case of the frequency comb heterodyning with ASE, we can simply multiply the single tooth PSD by the number of comb teeth $N$. The power spectral density is:
\begin{equation}
    \langle i(t)^2 \rangle_{\text{PSD,Comb}} = 4\frac{\eta^2 e^2}{(h \nu)^2} S P_t N = 4\frac{\eta^2 e^2}{(h \nu)^2} S P_c.
\end{equation}
Here $P_c$ is the total comb power. While this may make intuitive sense, we supply the following step-by-step derivation.

To move to the power spectral density of the comb heterodyne, we define the field of the comb as a sum over the $N$ comb teeth:
\begin{equation}
    E_{c}(t) = \sum_{n = 1}^N \sqrt{2 P_t} \cos(\omega_n t + \Theta_n),
\end{equation}
where $\Theta_n$ is an arbitrary phase. Each tooth is spaced $B_o/2$ away from the adjacent teeth, i.e., $\omega_1 - \omega_0 = f_r = B_o/2$ and follows the comb equation $\omega_n = n_c \times \omega_r + \omega_0 $. Across this bandwidth, the ASE can be represented as:
\begin{equation}
    E_{\text{ASE, tot}}(t) = \sum_{m = 1}^{N} \sum_{k = -B_o/4\delta \nu}^{B_o/4\delta \nu} \sqrt{2 S \delta \nu} \times \cos((\omega_{m} + 2 \pi k \delta \nu)t + \Phi_{k,m}),
\end{equation}
where $\Phi_{k,m}$ is a random phase. Note that our ASE bins are defined from $-B_o/4$ to $B_o/4$, or $-f_r/2$ to $f_r/2$. This choice of bin size counts the ASE exactly once. The photodetected power is then:
\begin{equation}
P_{\text{comb, ASE}}(t) = \bigg(\sum_{n = 1}^N \sqrt{2 P_t} \cos(\omega_n t + \Theta_n) + \sum_{m = 1}^{N} \sum_{k = -B_o/4\delta \nu}^{B_o/4\delta \nu} \sqrt{2 S \delta \nu} \times \cos((\omega_{m} + 2 \pi k \delta \nu)t + \Phi_{k,m}) \bigg)^2
\end{equation}
We expand this to:
\begin{multline}
P_{\text{comb, ASE}}(t) = \bigg(\sum_{n = 1}^N \sqrt{2 P_t} \cos(\omega_n t + \Theta_n)\bigg)^2 \\ + 2\sum_{n = 1}^N \sqrt{2 P_t} \cos(\omega_n t + \Theta_n) \times \sum_{m = 1}^N \sum_{k = -B_o/4\delta \nu}^{B_o/4\delta \nu} \sqrt{2 S \delta \nu} \times \cos((\omega_m + 2 \pi k \delta \nu)t + \Phi_{k,m}) \\ + \bigg(\sum_{m = 1}^N \sum_{k = -B_o/4\delta \nu}^{B_o/4\delta \nu} \sqrt{2 S \delta \nu} \times \cos((\omega_m + 2 \pi k \delta \nu)t + \Phi_{k,m}\bigg)^2.
\end{multline}
As before, only the heterodyne difference frequency terms remain after squaring:
\begin{equation}
P_{\text{comb, ASE, dif}}(t) = 2 \sqrt{2 P_t 2 S \delta \nu}  \sum_{n = 1}^N \sum_{m = 1}^N \sum_{k = -B_o/4\delta \nu}^{B_o/4\delta \nu} \frac{1}{2} \cos((\omega_n - \omega_m)t - 2 \pi k \delta \nu t + \Theta_n - \Phi_{k,m}).
\end{equation}
Given a detector RF bandwidth of $B_o/2 = f_r$, we note that a single $f_r$-sized bin of ASE does not heterodyne with all the comb teeth; instead, it heterodynes with the comb tooth that splits it and also the comb tooth at its upper edge and its lower edge. We can thus split the above power into three components:
\begin{multline}
P_{\text{comb, ASE, dif, three}}(t) = \sqrt{2 P_t 2 S \delta \nu}  \bigg(\sum_{n = 1, m = n}^N \sum_{k = -B_o/4\delta \nu}^{B_o/4\delta \nu} \cos((\omega_n - \omega_m)t - 2 \pi k \delta \nu t + \Theta_n - \Phi_{k,m}) \\ + \sum_{n = 1, m = n+1}^{N-1} \sum_{k = -B_o/4\delta \nu}^{0} \cos((\omega_n - \omega_m)t - 2 \pi k \delta \nu t + \Theta_n - \Phi_{k,m}) \\ + \sum_{n = 2, m = n-1}^N \sum_{k = 0}^{B_o/4\delta \nu} \cos((\omega_n - \omega_m)t - 2 \pi k \delta \nu t + \Theta_n - \Phi_{k,m}) \bigg).
\end{multline}
Next, we turn the power into a photocurrent:
\begin{multline}
i_{\text{comb,ASE}}(t) = \frac{\eta e}{h \nu} \sqrt{2 P_t 2 S \delta \nu}  \bigg(\sum_{n = 1, m = n}^N \sum_{k = -B_o/4\delta \nu}^{B_o/4\delta \nu} \cos((\omega_n - \omega_m)t - 2 \pi k \delta \nu t + \Theta_n - \Phi_{k,m}) \\ + \sum_{n = 1, m = n+1}^{N-1} \sum_{k = -B_o/4\delta \nu}^{0} \cos((\omega_n - \omega_m)t - 2 \pi k \delta \nu t + \Theta_n - \Phi_{k,m}) \\ + \sum_{n = 2, m = n-1}^N \sum_{k = 0}^{B_o/4\delta \nu} \cos((\omega_n - \omega_m)t - 2 \pi k \delta \nu t + \Theta_n - \Phi_{k,m}) \bigg).
\end{multline}
Because there are no correlated phases here, we take the root mean square of cosine and sum in quadrature:
\begin{equation}
\sqrt{\langle i_{\text{comb, ASE}}(t)^2 \rangle} = \frac{\eta e}{h \nu} \sqrt{2 P_t 2 S \delta \nu}  \frac{1}{\sqrt{2}}\bigg( \sqrt{N \frac{B_o}{2 \delta \nu}}^2 + \sqrt{(N-1) \frac{B_o}{4\delta \nu}}^2 + \sqrt{(N-1) \frac{B_o}{4\delta \nu}}^2\bigg)^\frac{1}{2}.
\end{equation}
Note that the first term comes from the central tooth heterodyne and is distributed as white noise from DC to $B_o/4$. The next two terms are from the outer tooth heterodynes and are distributed from $B_o/4$ to $B_o/2$. In fact, these terms add to form a white noise floor from DC to $B_o/2$. Simplifying the equation:
\begin{equation}
\sqrt{\langle i_{\text{comb, ASE}}(t)^2 \rangle} = \frac{\eta e}{h \nu} \sqrt{2 P_t 2 S \delta \nu}  \frac{1}{\sqrt{2}}\bigg( N \frac{B_o}{2\delta \nu} + (N-1) \frac{B_o}{2\delta \nu} \bigg)^\frac{1}{2}.
\end{equation}
Simplifying further:
\begin{equation}
\sqrt{\langle i_{\text{comb, ASE}}(t)^2 \rangle} = \frac{\eta e}{h \nu} \sqrt{2 P_t 2 S \delta \nu}  \frac{1}{\sqrt{2}} \sqrt{\frac{B_o}{2 \delta \nu} }\bigg( 2N - 1\bigg)^\frac{1}{2}.
\end{equation}
Further simplifying:
\begin{equation}
\sqrt{\langle i_{\text{comb, ASE}}(t)^2 \rangle} = \frac{\eta e}{h \nu} \sqrt{2 P_t 2 S \delta \nu}  \frac{1}{\sqrt{2}} \sqrt{\frac{B_o}{2 \delta \nu} }\sqrt{2 N - 1} .
\end{equation}
Canceling terms and simplifying:
\begin{equation}
\sqrt{\langle i_{\text{comb, ASE}}(t)^2 \rangle} = \frac{\eta e}{h \nu} \sqrt{ P_t  S B_o} \sqrt{2N-1} .
\end{equation}
The power is then:
\begin{equation}
\langle i_{\text{comb, ASE}}(t)^2 \rangle = \bigg(\frac{\eta e}{h \nu}\bigg)^2 P_t  S B_o (2N-1).
\end{equation}
We recover the PSD by dividing by the RF bandwidth $B_o/2$:
\begin{equation}
\langle i_{\text{comb, ASE}}(t)^2 \rangle_\text{PSD} = 2 \bigg(\frac{\eta e}{h \nu}\bigg)^2 P_t  S (2N-1).
\end{equation}
Now we make the assumption for large $N$, i.e., that $2N-1 \rightarrow 2N$:
\begin{equation}
\langle i_{\text{comb, ASE}}(t)^2 \rangle_\text{PSD} = 4 \bigg(\frac{\eta e}{h \nu}\bigg)^2 P_t  S N = 4 \bigg(\frac{\eta e}{h \nu}\bigg)^2 S P_c.
\end{equation}
This expression holds for both a single detector and balanced detector, assuming that the thermal power and comb power are the same in total. Consider the case of balanced detection:
\begin{equation}
\langle i_{\text{D1}}(t)^2 \rangle_\text{PSD} = \langle i_{\text{D2}}(t)^2 \rangle_\text{PSD} = 4 \bigg(\frac{\eta e}{h \nu}\bigg)^2 \frac{S}{2} \frac{P_c}{2} = \bigg(\frac{\eta e}{h \nu}\bigg)^2 S P_c,
\end{equation}
where these are the PSDs of the current outputs of the two detectors. Note that the thermal power and comb power to each detector is one half the total. Also note that this ``noise'' is correlated (identical), so the signal amplitudes add linearly, not in quadrature. I.e.:
\begin{equation}
\langle i_{\text{BPD}}(t)^2 \rangle_\text{PSD} = \Bigg(\sqrt{\bigg(\frac{\eta e}{h \nu}\bigg)^2 S P_c} + \sqrt{\bigg(\frac{\eta e}{h \nu}\bigg)^2 S P_c}\Bigg)^2 = 4 \bigg(\frac{\eta e}{h \nu}\bigg)^2 S P_c,
\end{equation}\label{thermalheterodynepsd}

\subsection{Simplified SNR Derivation}\label{SimpleSNRderivation}
We just derived that the electrical power spectral density of the heterodyne between thermal light and a frequency comb LO is the following, now rewritten with a subscript denoting thermal heterodyne and omitting the PSD subscript:
\begin{equation}
\langle i_T^2 \rangle = 4 \bigg(\frac{\eta e}{h \nu}\bigg)^2  S P_{LO},
\end{equation}
where $P_{LO} = P_c$. The shot noise PSD is:
\begin{equation}
\langle i_{SN}^2 \rangle = 2 \frac{\eta e^2}{h \nu} P_{LO}.
\end{equation}
The total root-mean-square current from the first balanced detector is $\sqrt{\langle i_{BD_1}^2 \rangle} = \sqrt{\langle i_{SN_1}^2 \rangle + \langle i_{T_1}^2 \rangle}$, and the total root-mean-square current from the other balanced detector is defined in the same way. The mixer multiplies these two photocurrents, such that the RMS output of the mixer is:
\begin{equation}
\sqrt{\langle i_{SN_1}^2 \rangle + \langle i_{T_1}^2 \rangle} \times \sqrt{\langle i_{SN_2}^2 \rangle + \langle i_{T_2}^2 \rangle}.
\end{equation}
The resulting terms are:

$\sqrt{\langle i_{SN_1}^2 \rangle \langle i_{SN_2}^2 \rangle}$, $\sqrt{\langle i_{SN_1}^2\rangle \langle i_{T_2}^2\rangle }$, $\sqrt{\langle i_{T_1}^2\rangle \langle i_{SN_2}^2\rangle }$, and $\sqrt{\langle i_{T_1}^2\rangle \langle i_{T_2}^2\rangle }$.

The last term defines the signal, and all terms, including the last, comprise the noise. Assuming the same levels of $S$ and $P_{LO}$ at each detector, the signal-to-noise ratio (SNR) is:
\begin{equation}
\text{SNR} = \frac{\langle i_T^2\rangle }{\sqrt{\langle i_T^2\rangle ^2 + \langle i_T^2\rangle \langle i_{SN}^2\rangle  + \langle i_T^2\rangle \langle i_{SN}^2\rangle  + \langle i_{SN}^2\rangle ^2}},
\end{equation}
where the mixer output terms are all added in quadrature since they are uncorrelated. Simplifying:
\begin{equation}
\text{SNR} = \frac{\langle i_T^2\rangle }{\sqrt{\langle i_T^2\rangle ^2 + 2\langle i_T^2\rangle \langle i_{SN}^2\rangle  + \langle i_{SN}^2\rangle ^2}} = \frac{\langle i_T^2\rangle }{\sqrt{[\langle i_T^2\rangle  + \langle i_{SN}^2\rangle ]^2}} .
\end{equation}
Further simplifying:
\begin{equation}
\text{SNR} = \frac{\langle i_T^2\rangle }{\langle i_T^2\rangle  + \langle i_{SN}^2\rangle } ,
\end{equation}
Plugging in the terms, where $S_{tot} = 2 S$, i.e., now looking at the total amount of thermal light on both balanced detectors:
\begin{equation}
\text{SNR} = \frac{2 \bigg(\frac{\eta e}{h \nu}\bigg)^2 P_{LO}  S_{tot}}{2 \bigg(\frac{\eta e}{h \nu}\bigg)^2 P_{LO}  S_{tot} + 2 \frac{\eta e^2}{h \nu} P_{LO}} = \frac{\frac{\eta}{h \nu}S_{tot}}{\frac{\eta}{h \nu}S_{tot} + 1} = \frac{\eta \langle n \rangle}{\eta \langle n \rangle + 1},
\end{equation}
where $ S_{tot} =  h \nu \langle n \rangle$. Note that there is some slight inconsistency with the main text, where $S$ refers to the total PSD. This notation is chosen for visual simplicity. However the appendices are themselves self-consistent.

\subsection{Frequency Domain Derivation of SNR}

Now we begin the more detailed derivation of SNR in the frequency domain, following a picture like that of Fig. 2.

\paragraph{Photodetection}
We begin by looking at the heterodyne beat photocurrent between a comb tooth and the thermal light right before the electronic mixer. We represent it over a bandwidth $2 B_f$ in terms of uncorrelated cosines. This follows techniques in Appendix \ref{CWhet}. 
\begin{equation}
   i_{\text{het, single tooth}} = \frac{\eta e}{h \nu} \sqrt{4 S_{det} \delta \nu P_t} \sum_{k = -B_f/\delta \nu}^{B_f/\delta \nu} \sqrt{2} \cos(2 \pi k \delta \nu t + \Phi_k - \Theta).
\end{equation}
$S_{det}$ is the power spectral density of the thermal light coming into each balanced detector. $B_f$ is the bandwidth of the RF low-pass filters we place after balanced detection. Note that this is an equivalent representation of the mean square value white noise of Equation \ref{thermalheterodynepsd}.

Next we represent the current arising from heterodynes of all comb teeth. We choose to have $N+1$ comb teeth. In other words, we have $N+1$ sums from the full comb interfering with the thermal light. This choice allows us to sum from $n = 0$ to $n=N$. Thus our full expression before the electronic mixer is:
\begin{equation}
   i_{\text{het}} = \frac{\eta e}{h \nu}  \sum_{n = 0}^{n = N} \sqrt{ 4 S_{det} \delta \nu P_t} \sum_{k = -B_f/\delta \nu}^{B_f/\delta \nu} \sqrt{2}\cos(2 \pi k \delta \nu t + \Phi_{k,n} - \Theta_n).
\end{equation}
We index the random phase ($\Phi_{k,n}$) by the comb number so that all cosine terms have random relative phases with respect to each other. We also index the comb tooth phase $\Theta_n$ by the comb tooth number since we are including many comb teeth now. We note that these teeth are spaced by $f_r$ in the frequency domain, where $f_r$ is the repetition rate. We assume that the total comb power is distributed evenly among all teeth (rectangular spectral profile).

Next, we consider the other comb and thermal signal. This is similar to the equation above:
\begin{equation}
    i_{\text{het}}^\prime = \frac{\eta e}{h \nu} \sum_{m = 0}^{m = N} \sqrt{ 4 S_{det} \delta \nu P_t^\prime} \sum_{j = -B_f/\delta \nu}^{B_f/\delta \nu} \sqrt{2} \cos(2 \pi j \delta \nu t+ \Phi_{j,m}^\prime - \Theta_m^\prime).
\end{equation}
We note two differences between this equation and the last equation. The second comb has a power per tooth of $P_t^\prime$, and the phases in the cosine argument are not the same, denoted by the prime notation. However, $\Phi_{k,n}$ and $\Phi_{j,m}^\prime$ have a special relationship when $n = m$; they are derived from the same quasi-thermal light. We thus have the relation:
\begin{equation}
    \Phi_{k,n} = \Phi_{j^\prime,n}^\prime
\end{equation}
when $j^\prime = k + \frac{n \Delta f_r}{\delta \nu}$. $\Delta f_r = f_r - f_r^\prime$, where $f_r^\prime$ is the repetition rate of the second comb and is assumed to be smaller than $f_r$.

At this point we should also take note of the shot noise from the first comb after balanced detection:
\begin{equation}
    i_{SN,RMS} = \sqrt{ 2 e i_{LO} B_f} = \sqrt{ 2 e \frac{\eta e}{h \nu} (N+1) P_t B_f}.
\end{equation}
Note that the total comb power is $(N+1)P_t$, i.e., the number of comb teeth multiplied by the power per tooth. In order to evaluate how the shot noise propagates through the electronics, we represent the shot noise current as a sum of cosines with uncorrelated phases.
\begin{multline}
i_{SN} = \sqrt{ \frac{2 e \frac{\eta e}{h \nu} (N+1) P_t B_f}{2 B_f (N+1)/\delta \nu}}  \sum_{n = 0}^N \sum_{k = -B_f/\delta \nu}^{B_f/\delta \nu} \sqrt{2} \cos(2 \pi k \delta \nu t + \phi_{k,n} ) \\ =  \sqrt{2e \frac{\eta e}{h \nu} P_t \delta \nu} \sum_{n = 0}^N \sum_{k = -B_f/\delta \nu}^{B_f/\delta \nu} \sqrt{2} \cos(2 \pi k \delta \nu t + \phi_{k,n} ) .
\end{multline}
Note that quadrature summation of the RMS values of the cosines takes $i_{SN}$ back to $i_{SN,RMS}$. Similarly, the RMS shot noise of the second comb is:
\begin{equation}
    i_{SN,RMS} = \sqrt{2 e i_{LO}^\prime B_f} = \sqrt{2 e \frac{\eta e}{h \nu} (N+1) P_t ^\prime B_f}.
\end{equation}
And the shot noise photocurrent of the second comb can be written as:
\begin{equation}
i_{SN}^\prime =  \sqrt{2 e \frac{\eta e}{h \nu} P_t^\prime \delta \nu} \sum_{m = 0}^N \sum_{k = -B_f/\delta \nu}^{B_f/\delta \nu} \sqrt{2} \cos(2 \pi j \delta \nu t + \phi_{j,m}^\prime ) .
\end{equation}
We may also have excess relative intensity noise (RIN) due to balancing that is not perfect. We can represent the RMS intensity noise current of the first comb as:
\begin{equation}
    i_{RIN,RMS} = \sqrt{(N+1) B_f S_{IN}}.
\end{equation}$S_{IN}$ is the power spectral density of the intensity noise photocurrent per comb tooth, in units of amps squared per root hertz. Likewise, for the other comb, the RMS intensity noise current is:
\begin{equation}
    i_{RIN,RMS}^\prime = \sqrt{(N+1) B_f S_{IN}^\prime}.
\end{equation}
It is most convenient for us to treat all the noise from the photodetectors at once, so we combine the shot noise and the intensity noise into a total photodetected noise term $i_{\text{PDN},RMS}$ for the first comb:
\begin{equation}
i_{\text{PDN},RMS} = \sqrt{i_{SN,RMS}^2 + i_{RIN,RMS}^2} = \sqrt{ \sqrt{2 e \frac{\eta e}{h \nu} (N+1) P_t B_f}^2 + \sqrt{(N+1) B_f S_{RIN}}^2}
\end{equation}
As before, we define this total photodetected noise in terms of a sum over cosines:
\begin{multline}
i_{\text{PDN}} = i_{\text{PDN},RMS} \sqrt{\frac{\delta \nu}{2 (N+1) B_f}} \sqrt{2}\sum_{n = 0}^N \sum_{k = -B_f/\delta \nu}^{B_f/\delta \nu} \cos(2 \pi k \delta \nu t + \phi_{k,n}^\prime ) \\ =  i_{\text{PDN},RMS} \sqrt{\frac{\delta \nu}{(N+1) B_f}} \sum_{n = 0}^N \sum_{k = -B_f/\delta \nu}^{B_f/\delta \nu} \cos(2 \pi k \delta \nu t + \phi_{k,n}^\prime )  .
\end{multline}
We define the total photodetected noise for the second comb similarly, such that:
\begin{equation}
i_{\text{PDN},RMS}^\prime = \sqrt{i_{SN,RMS}^{\prime 2} + i_{IN,RMS}^{\prime 2}} = \sqrt{ \sqrt{2 e \frac{\eta e}{h \nu} (N+1) P_t^\prime B_f}^2 + \sqrt{(N+1) B_f S_{RIN}^\prime}^2},
\end{equation}
and:
\begin{equation}
i_{\text{PDN}}^\prime =  i_{\text{PDN},RMS}^\prime \sqrt{\frac{\delta \nu}{(N+1) B_f}} \sum_{m = 0}^N \sum_{j = -B_f/\delta \nu}^{B_f/\delta \nu} \cos(2 \pi j \delta \nu t + \phi_{j,m}^\prime )  .
\end{equation}
Thus, the total current of the first comb heterodyne signal before mixing is:
\begin{equation}
    i_{\text{tot}} = i_{\text{het}} + i_{\text{PDN}},
\end{equation}
and the second is:
\begin{equation}
    i_{\text{tot}}^\prime = i_{\text{het}} ^\prime + i_{\text{PDN}}^\prime.
\end{equation}
If limited by light noise, these are the currents that result from the two detectors. Next we move to the mixing stage.

\paragraph{Mixing of Heterodyne Signal with Heterodyne Signal}

Mixing the two heterodyne signals results in:
\begin{equation}
    i_{\text{mix}} = i_{\text{tot}} \times i_\text{tot}^\prime = (i_\text{het} + i_{\text{PDN}}) \times (i_\text{het} ^\prime + i_{\text{PDN}}^\prime) = i_\text{het} i_\text{het}^\prime + i_\text{het} i_{\text{PDN}}^\prime + i_{\text{PDN}} i_\text{het}^\prime + i_{\text{PDN}} i_{\text{PDN}}^\prime.
\end{equation}
We will see that the first term that is the product of the two heterodyne currents contributes to both noise and signal, while the three other terms contribute to noise. First we examine the thermal heterodyne product term.
\begin{multline}
    i_\text{het}i_\text{het} ^\prime = \frac{\eta e}{h \nu}  \sum_{n = 0}^{n = N} \sqrt{4 S_{det} \delta \nu P_t} \sum_{k = -B_f/\delta \nu}^{B_f/\delta \nu} \sqrt{2} \cos(2 \pi k \delta \nu t + \Phi_{k,n} - \Theta_n) \\
    \times \frac{\eta e}{h \nu} \sum_{m = 0}^{m = N} \sqrt{4 S_{det} \delta \nu P_t^\prime} \sum_{j = -B_f/\delta \nu}^{B_f/\delta \nu} \sqrt{2} \cos(2 \pi j \delta \nu t + \Phi_{j,m}^\prime - \Theta_m^\prime)
\end{multline}
Simplifying, we have:
\begin{multline}
    i_\text{het}i_\text{het} ^\prime = (\frac{\eta e}{h \nu})^2 8 (S_{det} \delta \nu) \sqrt{P_t P_t^\prime}  \\ \times \sum_{n = 0}^{n = N} \sum_{m = 0}^{m = N} \sum_{k = -B_f/\delta \nu}^{B_f/\delta \nu} \sum_{j = -B_f/\delta \nu}^{B_f/\delta \nu} \cos(2 \pi k \delta \nu t + \Phi_{k,n} - \Theta_n) \cos(2 \pi j \delta \nu t + \Phi_{j,m}^\prime - \Theta_m^\prime)
\end{multline}
At this point, we note that the quasi-thermal light is uncorrelated across its spectrum. Thus, when $n \neq m$, $\Phi_{k,n}$ and $\Phi_{j,m}^\prime$ are uncorrelated regardless of $k$ and $j$. However, when $n = m$, $\Phi_{k,n}$ and $\Phi_{j,m}^\prime$ are correlated (i.e., the same) under the condition that $j = k + \frac{n \Delta f_r}{\delta \nu}$. We then split the above heterodyne product into two terms: when $n \neq m$ and when $n = m$.
\begin{multline}
    i_\text{het}i_\text{het} ^\prime = (\frac{\eta e}{h \nu})^2 8 (S_{det} \delta \nu) \sqrt{P_t P_t^\prime}  \\
    \times \bigg( \sum_{n = 0}^{n = N} \sum_{m = 0, m \neq n}^{m = N} \sum_{k = -B_f/\delta \nu}^{B_f/\delta \nu} \sum_{j = -B_f/\delta \nu}^{B_f/\delta \nu} \cos(2 \pi k \delta \nu t + \Phi_{k,n} - \Theta_n) \cos(2 \pi j \delta \nu t + \Phi_{j,m}^\prime - \Theta_m^\prime) \\
    + \sum_{n = 0, n = m}^{n = N} \sum_{k = -B_f/\delta \nu}^{B_f/\delta \nu} \sum_{j = -B_f/\delta \nu}^{B_f/\delta \nu} \cos(2 \pi k \delta \nu t + \Phi_{k,n} - \Theta_n) \cos(2 \pi j \delta \nu t + \Phi_{j,m}^\prime - \Theta_m^\prime) \bigg)
\end{multline}
We must further break down the $n = m$ product into those without correlated phases and those with correlated phases (i.e., split the sum):
\begin{multline}
    i_\text{het}i_\text{het} ^\prime = (\frac{\eta e}{h \nu})^2 8 (S_{det} \delta \nu) \sqrt{P_t P_t^\prime}  \\
    \times \bigg( \sum_{n = 0}^{n = N} \sum_{m = 0, m \neq n}^{m = N} \sum_{k = -B_f/\delta \nu}^{B_f/\delta \nu} \sum_{j = -B_f/\delta \nu}^{B_f/\delta \nu} \cos(2 \pi k \delta \nu t + \Phi_{k,n} - \Theta_n) \cos(2 \pi j \delta \nu t + \Phi_{j,m}^\prime - \Theta_m^\prime) \\
    + \sum_{n = 0, n = m}^{n = N} \sum_{k = -B_f/\delta \nu}^{B_f/\delta \nu} \sum_{j = -B_f/\delta \nu, j \neq k + \frac{n \Delta f_r}{\delta \nu}}^{B_f/\delta \nu} \cos(2 \pi k \delta \nu t + \Phi_{k,n} - \Theta_n) \cos(2 \pi j \delta \nu t + \Phi_{j,m}^\prime - \Theta_m^\prime) \\
    + \sum_{n = 0, n = m}^{n = N} \sum_{j = -B_f/\delta \nu, j = k + \frac{n \Delta f_r}{\delta \nu}}^{B_f/\delta \nu} \cos(2 \pi k \delta \nu t + \Phi_{k,n} - \Theta_n) \cos(2 \pi j \delta \nu t + \Phi_{j,m}^\prime - \Theta_m^\prime) \bigg)
\end{multline}
However, even here in the last term there exist products that do not contain correlated phases. These uncorrelated product terms are the ones residing at the outer edges of the bandwidth $B_f$. To fully capture this, we must split the last term into three more terms: one term from the middle section and two terms from the edge of the bandwidth to $n\Delta f_r$ inside.
\begin{multline}
    i_\text{het}i_\text{het} ^\prime = (\frac{\eta e}{h \nu})^2 8 (S_{det} \delta \nu) \sqrt{P_t P_t^\prime}  \\
    \times \bigg( \sum_{n = 0}^{n = N} \sum_{m = 0, m \neq n}^{m = N} \sum_{k = -B_f/\delta \nu}^{B_f/\delta \nu} \sum_{j = -B_f/\delta \nu}^{B_f/\delta \nu} \cos(2 \pi k \delta \nu t + \Phi_{k,n} - \Theta_n) \cos(2 \pi j \delta \nu t + \Phi_{j,m}^\prime - \Theta_m^\prime) \\
    + \sum_{n = 0, n = m}^{n = N} \sum_{k = -B_f/\delta \nu}^{B_f/\delta \nu} \sum_{j = -B_f/\delta \nu, j \neq k + \frac{n \Delta f_r}{\delta \nu}}^{B_f/\delta \nu} \cos(2 \pi k \delta \nu t + \Phi_{k,n} - \Theta_n) \cos(2 \pi j \delta \nu t + \Phi_{j,m}^\prime - \Theta_m^\prime) \\
    + \sum_{n = 0, n = m}^{n = N} \sum_{j = -B_f/\delta \nu, j = k + \frac{n \Delta f_r}{\delta \nu}}^{-B_f/\delta \nu + \frac{n \Delta f_r}{\delta \nu}} \cos(2 \pi k \delta \nu t + \Phi_{k,n} - \Theta_n) \cos(2 \pi j \delta \nu t + \Phi_{j,m}^\prime - \Theta_m^\prime) \\
    + \sum_{n = 0, n = m}^{n = N} \sum_{j = B_f/\delta \nu - \frac{n \Delta f_r}{\delta \nu}, j = k + \frac{n \Delta f_r}{\delta \nu}}^{B_f/\delta \nu} \cos(2 \pi k \delta \nu t + \Phi_{k,n} - \Theta_n) \cos(2 \pi j \delta \nu t + \Phi_{j,m}^\prime - \Theta_m^\prime) \\
    + \sum_{n = 0, n = m}^{n = N} \sum_{j = -B_f/\delta \nu + \frac{n \Delta f_r}{\delta \nu}, j = k + \frac{n \Delta f_r}{\delta \nu}}^{B_f/\delta \nu - \frac{n \Delta f_r}{\delta \nu}} \cos(2 \pi k \delta \nu t + \Phi_{k,n} - \Theta_n) \cos(2 \pi j \delta \nu t + \Phi_{j,m}^\prime - \Theta_m^\prime) \bigg)
\end{multline} \label{eqn: hetmixfull}
Next, we simplify the cosine products, identical for each sum.
\begin{multline}
    \cos(2 \pi k \delta \nu + \Phi_{k,n} - \Theta_n) \cos(2 \pi j \delta \nu + \Phi_{j,m}^\prime - \Theta_m^\prime) \\
    = \frac{\cos(2 \pi (k+j) \delta \nu t + \Phi_{k,n} - \Theta_n  + \Phi_{j,m}^\prime - \Theta_m^\prime)}{2} + \frac{\cos(2 \pi (k-j) \delta \nu t + \Phi_{k,n} - \Theta_n - \Phi_{j,m}^\prime + \Theta_m^\prime)}{2}
\end{multline}
We substitute the cosine products into the mixed heterodyne signal equation:
\begin{multline}
    i_\text{het}i_\text{het} ^\prime = (\frac{\eta e}{h \nu})^2 8 (S_{det} \delta \nu) \sqrt{P_t P_t^\prime}  \\
    \times \bigg( \sum_{n = 0}^{n = N} \sum_{m = 0, m \neq n}^{m = N} \sum_{k = -B_f/\delta \nu}^{B_f/\delta \nu} \sum_{j = -B_f/\delta \nu}^{B_f/\delta \nu} \\
    \frac{\cos(2 \pi (k+j) \delta \nu t + \Phi_{k,n} - \Theta_n  + \Phi_{j,m}^\prime - \Theta_m^\prime)}{2} + \frac{\cos(2 \pi (k-j) \delta \nu t + \Phi_{k,n} - \Theta_n - \Phi_{j,m}^\prime + \Theta_m^\prime)}{2} \\
    + \sum_{n = 0, n = m}^{n = N} \sum_{k = -B_f/\delta \nu}^{B_f/\delta \nu} \sum_{j = -B_f/\delta \nu, j \neq k + \frac{n \Delta f_r}{\delta \nu}}^{B_f/\delta \nu} \\
    \frac{\cos(2 \pi (k+j) \delta \nu t + \Phi_{k,n} - \Theta_n  + \Phi_{j,m}^\prime - \Theta_m^\prime)}{2} + \frac{\cos(2 \pi (k-j) \delta \nu t + \Phi_{k,n} - \Theta_n - \Phi_{j,m}^\prime + \Theta_m^\prime)}{2} \\
    + \sum_{n = 0, n = m}^{n = N} \sum_{j = -B_f/\delta \nu, j = k + \frac{n \Delta f_r}{\delta \nu}}^{-B_f/\delta \nu + \frac{n \Delta f_r}{\delta \nu}} \\
    \frac{\cos(2 \pi (k+j) \delta \nu t + \Phi_{k,n} - \Theta_n  + \Phi_{j,m}^\prime - \Theta_m^\prime)}{2} + \frac{\cos(2 \pi (k-j) \delta \nu t + \Phi_{k,n} - \Theta_n - \Phi_{j,m}^\prime + \Theta_m^\prime)}{2} \\
    + \sum_{n = 0, n = m}^{n = N} \sum_{j = B_f/\delta \nu - \frac{n \Delta f_r}{\delta \nu}, j = k + \frac{n \Delta f_r}{\delta \nu}}^{B_f/\delta \nu} \\
    \frac{\cos(2 \pi (k+j) \delta \nu t + \Phi_{k,n} - \Theta_n  + \Phi_{j,m}^\prime - \Theta_m^\prime)}{2} + \frac{\cos(2 \pi (k-j) \delta \nu t + \Phi_{k,n} - \Theta_n - \Phi_{j,m}^\prime + \Theta_m^\prime)}{2} \\
    + \sum_{n = 0, n = m}^{n = N} \sum_{j = -B_f/\delta \nu + \frac{n \Delta f_r}{\delta \nu}, j = k + \frac{n \Delta f_r}{\delta \nu}}^{B_f/\delta \nu - \frac{n \Delta f_r}{\delta \nu}} \\
    \frac{\cos(2 \pi (k+j) \delta \nu t + \Phi_{k,n} - \Theta_n  + \Phi_{j,m}^\prime - \Theta_m^\prime)}{2} + \frac{\cos(2 \pi (k-j) \delta \nu t + \Phi_{k,n} - \Theta_n - \Phi_{j,m}^\prime + \Theta_m^\prime)}{2} \bigg) \\
\end{multline}
These correspond to the currents directly after the mixer. After the mixer, we now consider a low-pass filter $B_{f,m}$ to block the high frequency noise. In the simplest case, we choose a filter where $B_{f,m} = B_f$. After the mixer, the mixed heterodyne signals are the following, where only the sum terms only cover $|k+j| < B_{f,m}/
\delta \nu$, and the difference terms only cover $|k-j| < B_{f,m}/
\delta \nu$ due to the low pass filter bandwidth of $B_{f,m}$:
\begin{multline}
\begin{aligned}
    i_\text{het}i_\text{het} ^\prime &= (\frac{\eta e}{h \nu})^2 8 (S_{det} \delta \nu) \sqrt{P_t P_t^\prime}  \\
    &\times \bigg( \sum_{n = 0}^{n = N} \sum_{m = 0, m \neq n}^{m = N} \sum_{k = -B_f/\delta \nu}^{B_f/\delta \nu} \sum_{j = -B_f/\delta \nu, |k + j| < B_{f,m}/\delta \nu }^{B_f/\delta \nu}
    \!\!\!\!\!\!\!\!\!\!\!\!\!\!\!\!\!\!\!\!\frac{\cos(2 \pi (k+j) \delta \nu t + \Phi_{k,n} - \Theta_n  + \Phi_{j,m}^\prime - \Theta_m^\prime)}{2} \\
    &+  \sum_{n = 0}^{n = N} \sum_{m = 0, m \neq n}^{m = N} \sum_{k = -B_f/\delta \nu}^{B_f/\delta \nu} \sum_{j = -B_f/\delta \nu, |k - j| < B_{f,m}/\delta \nu}^{B_f/\delta \nu} \frac{\cos(2 \pi (k-j) \delta \nu t + \Phi_{k,n} - \Theta_n - \Phi_{j,m}^\prime + \Theta_m^\prime)}{2} \\
    &+ \sum_{n = 0, n = m}^{n = N} \sum_{k = -B_f/\delta \nu}^{B_f/\delta \nu} \sum_{j = -B_f/\delta \nu, j \neq k + \frac{n \Delta f_r}{\delta \nu}, |k + j| < B_{f,m}/\delta \nu}^{B_f/\delta \nu} \frac{\cos(2 \pi (k+j) \delta \nu t + \Phi_{k,n} - \Theta_n  + \Phi_{j,m}^\prime - \Theta_m^\prime)}{2} \\
    &+ \sum_{n = 0, n = m}^{n = N} \sum_{k = -B_f/\delta \nu}^{B_f/\delta \nu} \sum_{j = -B_f/\delta \nu, j \neq k + \frac{n \Delta f_r}{\delta \nu},|k - j| < B_{f,m}/\delta \nu}^{B_f/\delta \nu} \frac{\cos(2 \pi (k-j) \delta \nu t + \Phi_{k,n} - \Theta_n - \Phi_{j,m}^\prime + \Theta_m^\prime)}{2} \\
    &+ \sum_{n = 0, n = m}^{n = N} \sum_{j = -B_f/\delta \nu, j = k + \frac{n \Delta f_r}{\delta \nu},|k + j| < B_{f,m}/\delta \nu}^{-B_f/\delta \nu + \frac{n \Delta f_r}{\delta \nu}}
    \frac{\cos(2 \pi (k+j) \delta \nu t + \Phi_{k,n} - \Theta_n  + \Phi_{j,m}^\prime - \Theta_m^\prime)}{2} \\
    &+ \sum_{n = 0, n = m}^{n = N} \sum_{j = -B_f/\delta \nu, j = k + \frac{n \Delta f_r}{\delta \nu},|k - j| < B_{f,m}/\delta \nu}^{-B_f/\delta \nu + \frac{n \Delta f_r}{\delta \nu}} \frac{\cos(2 \pi (k-j) \delta \nu t + \Phi_{k,n} - \Theta_n - \Phi_{j,m}^\prime + \Theta_m^\prime)}{2} \\
    &+ \sum_{n = 0, n = m}^{n = N} \sum_{j = B_f/\delta \nu - \frac{n \Delta f_r}{\delta \nu}, j = k + \frac{n \Delta f_r}{\delta \nu},|k + j| < B_{f,m}/\delta \nu}^{B_f/\delta \nu}
    \frac{\cos(2 \pi (k+j) \delta \nu t + \Phi_{k,n} - \Theta_n  + \Phi_{j,m}^\prime - \Theta_m^\prime)}{2} \\
    &+ \sum_{n = 0, n = m}^{n = N} \sum_{j = B_f/\delta \nu - \frac{n \Delta f_r}{\delta \nu}, j = k + \frac{n \Delta f_r}{\delta \nu},|k - j| < B_{f,m}/\delta \nu}^{B_f/\delta \nu}  \frac{\cos(2 \pi (k-j) \delta \nu t + \Phi_{k,n} - \Theta_n - \Phi_{j,m}^\prime + \Theta_m^\prime)}{2} \\
    &+ \sum_{n = 0, n = m}^{n = N} \sum_{j = -B_f/\delta \nu + \frac{n \Delta f_r}{\delta \nu}, j = k + \frac{n \Delta f_r}{\delta \nu},|k + j| < B_{f,m}/\delta \nu}^{B_f/\delta \nu - \frac{n \Delta f_r}{\delta \nu}}
    \frac{\cos(2 \pi (k+j) \delta \nu t + \Phi_{k,n} - \Theta_n  + \Phi_{j,m}^\prime - \Theta_m^\prime)}{2} \\
    &+ \sum_{n = 0, n = m}^{n = N} \sum_{j = -B_f/\delta \nu + \frac{n \Delta f_r}{\delta \nu}, j = k + \frac{n \Delta f_r}{\delta \nu},|k - j| < B_{f,m}/\delta \nu}^{B_f/\delta \nu - \frac{n \Delta f_r}{\delta \nu}} \frac{\cos(2 \pi (k-j) \delta \nu t + \Phi_{k,n} - \Theta_n - \Phi_{j,m}^\prime + \Theta_m^\prime)}{2} \bigg) \\
    \end{aligned}
\end{multline} \label{eqn:fullheterodyneterms}
At this point, we will go through and describe each term again:
\begin{itemize}
    \item The first term contains sum frequency terms from non-same bins (i.e., no spectral overlap of the quasi-thermal light).
    
    \item The second term contains the difference frequency terms of non-same bins.
    
    \item The third term contains sum frequency terms of matching bins, but between intermediate frequencies that are not correlated. 
    \item The fourth term contains difference frequency terms of matching bins, but between intermediate frequencies that are not correlated. 
    \item The fifth and seventh terms contain sum frequency terms of matching bins that are not correlated due to the offset of the two teeth from separate combs--these sums correspond to the lower and upper non-overlapped optical regions. 
    \item Likewise, the sixth and eighth terms contain difference frequency terms of matching bins that are not correlated due to the offset of the two teeth from separate combs--again, these sums correspond to the lower and upper non-overlapped optical regions. 
    \item The ninth (second to last) term contains sum frequency terms between correlated heterodyne terms--however, there is no cancellation of phase due to sum frequency addition, so these terms are, like all those preceding it, uncorrelated. \item The tenth and final term contains difference frequency terms between correlated heterodyne terms; \textbf{this contains our signal.}
\end{itemize}

We separate this tenth term, our mixed heterodyne signal, into two parts, the noise and the signal, where $i_\text{het} i_\text{het}^\prime = N_\text{het} + A_\text{het}$ such that:
\begin{multline}
    A_\text{het} = (\frac{\eta e}{h \nu})^2 8 (S_{det} \delta \nu) \sqrt{P_t P_t^\prime} \\ \sum_{n = 0, n = m}^{n = N} \sum_{j = -B_f/\delta \nu + \frac{n \Delta f_r}{\delta \nu}, j = k + \frac{n \Delta f_r}{\delta \nu},|k - j| < B_{f,m}/\delta \nu}^{B_f/\delta \nu - \frac{n \Delta f_r}{\delta \nu}} \frac{\cos(2 \pi (k-j) \delta \nu t + \Phi_{k,n} - \Theta_n - \Phi_{j,m}^\prime + \Theta_m^\prime)}{2}
\end{multline}
Because we examine interferograms in this method, we will look at the peak value of the signal in the time domain. The maximal value occurs when we assume that both $\Theta_n$ and $\Theta_m^\prime$ are linear across the optical frequencies spanned by $N$ comb teeth. This linearity ensures constructive addition at the center of the resulting interferogram. Note that we also consider the low pass filter of $B_{f,m} = B_f$, which filters out 3/4 of the heterodyne tones. This is because the frequencies of this mixed term follow a triangular distribution centered at 0 with a width of $4B_f$; filtering at $\pm B_f$ cuts out 1/4 of these tones.  The phases $\Phi_{k,n}$ and $\Phi_{j,m}^\prime$ cancel at the indicated indices. We also sum over the $N$ comb teeth, resulting in $N+1$ bins. Thus our peak signal is:
\begin{equation}
    A_\text{het,peak} = (\frac{\eta e}{h \nu})^2 8 (S_{det} \delta \nu) \sqrt{P_t P_t^\prime} \times (N+1) \times  \frac{3}{4} \frac{2 B_f}{\delta \nu} \frac{1}{2}
\end{equation}
Simplifying,
\begin{equation}
    A_\text{het,peak} = 8 \times \frac{3}{4}(\frac{\eta e}{h \nu})^2 S_{det} \sqrt{P_t P_t^\prime} \times (N+1) \times B_f
\end{equation}

Next, we take the RMS of the heterodyne noise terms, i.e., all but the last term in Eqn. \ref{eqn:fullheterodyneterms} as follows.
\begin{multline}
     N_\text{het,rms} = (\frac{\eta e}{h \nu})^2 8 (S_{det} \delta \nu) \sqrt{P_t P_t^\prime}  \\
    \times 
    \bigg( (\sqrt{N(N+1)} \sqrt{\frac{3}{4} \frac{2 B_f}{ \delta \nu} \frac{3}{4}\frac{2 B_f}{ \delta \nu}} \frac{1}{2\sqrt{2}})^2 
    + ( \sqrt{N(N+1)} \sqrt{\frac{3}{4}\frac{2 B_f}{ \delta \nu} \frac{3}{4}\frac{2 B_f}{ \delta \nu}} \frac{1}{2\sqrt{2}})^2 \\
    + (\sqrt{N+1} \sqrt{\frac{3}{4}\frac{2 B_f}{ \delta \nu} \frac{3}{4}(\frac{2 B_f}{ \delta \nu}-1)} \frac{1}{2\sqrt{2}})^2 
    + (\sqrt{N+1} \sqrt{\frac{3}{4}\frac{2 B_f}{ \delta \nu} \frac{3}{4} (\frac{2 B_f}{ \delta \nu}-1)} \frac{1}{2\sqrt{2}})^2 \\
    + (\sqrt{N+1} \sqrt{\frac{3}{4}\frac{(2 B_f - \frac{n \Delta f_r}{\delta \nu})}{ \delta \nu}} \frac{1}{2\sqrt{2}})^2 + (\sqrt{N+1} \sqrt{\frac{3}{4}\frac{(2 B_f - \frac{n \Delta f_r}{\delta \nu})}{ \delta \nu}} \frac{1}{2\sqrt{2}})^2  \\ 
    + (\sqrt{N+1} \sqrt{\frac{3}{4}\frac{(\frac{n \Delta f_r}{\delta \nu})}{ \delta \nu}} \frac{1}{2\sqrt{2}})^2 
    + (\sqrt{N+1} \sqrt{\frac{3}{4}\frac{(\frac{n \Delta f_r}{\delta \nu})}{ \delta \nu}} \frac{1}{2\sqrt{2}})^2    
    + (\sqrt{N+1} \sqrt{\frac{3}{4}\frac{2 B_f }{ \delta \nu}} \frac{1}{2\sqrt{2}})^2 \bigg)^\frac{1}{2}
\end{multline}

Interestingly, these terms all follow a nested structure. We could simplify further and find that the nesting in fact makes the terms whole, but here we make size arguments to simplify. The first two terms are much larger than the following two terms by a factor of N, and the following two terms are a factor $\frac{B_f}{\delta \nu} $ much larger than the terms that follow those. We assume that $\frac{B_f}{\delta \nu} \gg 1$, thus for the first four terms we have:
\begin{multline}
     N_{\text{het}, RMS} = (\frac{\eta e}{h \nu})^2 8 (S_{det} \delta \nu) \sqrt{P_t P_t^\prime} \frac{1}{2\sqrt{2}}\\
    \times \bigg( N(N+1) \frac{3}{4} \frac{2 B_f}{ \delta \nu}\frac{3}{4} \frac{2 B_f}{ \delta \nu} +  N(N+1) \frac{3}{4} \frac{2 B_f}{ \delta \nu} \frac{3}{4}\frac{2 B_f}{ \delta \nu}
    + (N+1) \frac{3}{4}\frac{2 B_f}{ \delta \nu} \frac{3}{4}\frac{2 B_f}{ \delta \nu} + (N+1) \frac{3}{4} \frac{2 B_f}{ \delta \nu} \frac{3}{4}\frac{2 B_f}{ \delta \nu} \bigg)^\frac{1}{2}
\end{multline}
We can further combine terms:
\begin{equation}
     N_{\text{het}, RMS} = (\frac{\eta e}{h \nu})^2 8 (S_{det} \delta \nu) \sqrt{P_t P_t^\prime} \frac{1}{2\sqrt{2}}
    \times \frac{3 \sqrt{2}}{4} \bigg( (N+1)^2 \frac{2 B_f}{ \delta \nu} \frac{2 B_f}{ \delta \nu} \bigg)^\frac{1}{2}
\end{equation}
Taking the square root and simplifying, we have:
\begin{equation}
     N_{\text{het}, RMS} = 8 \times \frac{3}{4} (\frac{\eta e}{h \nu})^2 S_{det} \sqrt{P_t P_t^\prime} \times
    (N+1) \times B_f
\end{equation}
\paragraph{Mixing of Heterodyne Noise and Photodetected Noise}

Now, we examine another source of noise: the mixing of the photocurrents from the optical heterodyne and the photodetected noise.

\begin{multline}
N_{\text{PDN},\text{het}} = \frac{\eta e}{h \nu}  \sum_{n = 0}^{n = N} \sqrt{4 S_{det} \delta \nu P_t} \sum_{k = -B_f/\delta \nu}^{B_f/\delta \nu} \cos(2 \pi k \delta \nu t + \Phi_{k,n} - \Theta_n) \\
\times i_{\text{PDN},RMS}^\prime \sqrt{\frac{\delta \nu}{(N+1) B_f}} \sum_{m = 0}^N \sum_{j = -B_f/\delta \nu}^{B_f/\delta \nu} \cos(2 \pi j \delta \nu t + \phi_{j,m}^\prime )  .
\end{multline}
We simplify and combine terms:
\begin{multline}
N_{\text{PDN},\text{het}} = \frac{\eta e}{h \nu} \sqrt{4 S_{det} \delta \nu P_t} i_{\text{PDN},RMS}^\prime \sqrt{\frac{\delta \nu}{(N+1) B_f}} \\ \times \sum_{n = 0}^{n = N} \sum_{k = -B_f/\delta \nu}^{B_f/\delta \nu} \sum_{m = 0}^N \sum_{j = -B_f/\delta \nu}^{B_f/\delta \nu} \cos(2 \pi k \delta \nu t + \Phi_{k,n} - \Theta_n)\cos(2 \pi j \delta \nu t + \phi_{j,m}^\prime ) 
\end{multline}
Like our treatment of the mixed optical heterodyne signals, we separate the cosine product into sum and difference components. This noise also passes through the same $B_{f,m}$ bandpass filter after the mixer, thus constraining the number of included sum frequency components.
\begin{multline}
N_{\text{\text{PDN}},\text{het}} = \frac{\eta e}{h \nu} \sqrt{4 S_{det} \delta \nu P_t} i_{\text{PDN},RMS}^\prime \sqrt{\frac{\delta \nu}{(N+1) B_f}} \\ \times \Bigg( \sum_{n = 0}^{n = N} \sum_{k = -B_f/\delta \nu}^{B_f/\delta \nu} \sum_{m = 0}^N \sum_{j = -B_f/\delta \nu}^{B_f/\delta \nu} \cos(2 \pi k \delta \nu t + \Phi_{k,n} - \Theta_n - (2 \pi j \delta \nu t + \phi_{j,m}^\prime)) \\
+ \sum_{n = 0}^{n = N} \sum_{k = -B_f/\delta \nu}^{B_f/\delta \nu} \sum_{m = 0}^N \sum_{j = -B_f/\delta \nu, |j+k| < B_{f,m}/\delta \nu}^{B_f/\delta \nu} \cos(2 \pi k \delta \nu t + \Phi_{k,n} - \Theta_n + 2 \pi j \delta \nu t + \phi_{j,m}^\prime) \Bigg).
\end{multline}
Again, choosing $B_{f,m} = B_f$, the number of terms is multiplied by 3/4. Taking the RMS of the cosines and summing in quadrature, we have:
\begin{multline}
N_{\text{\text{PDN}},\text{het},RMS} = \frac{\eta e}{h \nu} \sqrt{4 S_{det} \delta \nu P_t} i_{\text{\text{PDN}},RMS}^\prime \sqrt{\frac{\delta \nu}{(N+1) B_f}} \\ \times \sqrt{ \bigg(\frac{3}{4} \sqrt{(N+1)^2 \bigg(\frac{2 B_f}{\delta \nu} \bigg)^2} \frac{1}{\sqrt{2}} \bigg)^2 + \bigg( \frac{3}{4} \sqrt{ (N+1)^2 \bigg(\frac{2 B_f}{\delta \nu}\bigg)^2} \frac{1}{\sqrt{2}}\bigg)^2 }.
\end{multline}
Simplifying the square root:
\begin{multline}
N_{\text{\text{PDN}},\text{het},RMS} = \frac{\eta e}{h \nu} \sqrt{4 S_{det} \delta \nu P_t} i_{\text{\text{PDN}},RMS}^\prime \sqrt{\frac{\delta \nu}{(N+1) B_f}} \\ \times \frac{1}{\sqrt{2}} \sqrt{ \bigg(\frac{3}{4}\bigg)^2 (N+1)^2 \bigg(\frac{2 B_f}{\delta \nu} \bigg)^2 + \bigg(\frac{3}{4}\bigg)^2(N+1)^2 \bigg(\frac{2 B_f}{\delta \nu} \bigg)^2}.
\end{multline}
Further simplifying:
\begin{multline}
N_{\text{\text{PDN}},\text{het},RMS} = \frac{\eta e}{h \nu} \sqrt{4 S_{det} \delta \nu P_t} i_{\text{\text{PDN}},RMS}^\prime \sqrt{\frac{\delta \nu}{(N+1) B_f}} \\ \times \frac{\sqrt{2}}{\sqrt{2}} \frac{3}{4} \sqrt{ (N+1)^2 \bigg( \frac{2 B_f}{\delta \nu} \bigg)^2 }.
\end{multline}
Simplifying the square root again:
\begin{equation}
N_{\text{\text{PDN}},\text{het},RMS} = \frac{\eta e}{h \nu} \sqrt{4 S_{det} \delta \nu P_t} i_{\text{\text{PDN}},RMS}^\prime \sqrt{\frac{\delta \nu}{(N+1) B_f}} \times \frac{3}{4}  (N+1) \frac{2 B_f}{\delta \nu}.
\end{equation}
Simplifying:
\begin{equation}
N_{\text{\text{PDN}},\text{het},RMS} = 2\times \frac{3}{4} \frac{\eta e}{h \nu} \sqrt{4 S_{det} P_t} i_{\text{\text{PDN}},RMS}^\prime \sqrt{(N+1) B_f} .
\end{equation}
By the same argument, the noise from the other sum is:
\begin{equation}
N_{\text{\text{PDN}},\text{het},RMS}^\prime = 2\times \frac{3}{4} \frac{\eta e}{h \nu} \sqrt{4 S_{det} P_t^\prime} i_{\text{\text{PDN}},RMS} \sqrt{(N+1) B_f} .
\end{equation}

\paragraph{Mixing of Photodetected Noise and Photodetected Noise}

Next, we look to the last noise term, when the photodetected noise from one comb mixes with the photodetected noise from the other comb.
\begin{multline}
    N_{\text{\text{PDN}},\text{PDN}} = i_{\text{PDN},\text{RMS}} \sqrt{\frac{\delta \nu}{(N+1) B_f}} \sum_{n = 0}^N \sum_{k = -B_f/\delta \nu}^{B_f/\delta \nu} \cos(2 \pi k \delta \nu t + \phi_{k,n} ) \\ \times i_{\text{PDN},\text{RMS}}^\prime \sqrt{\frac{\delta \nu}{(N+1) B_f}} \sum_{m = 0}^N \sum_{j = -B_f/\delta \nu}^{B_f/\delta \nu} \cos(2 \pi j \delta \nu t + \phi_{j,m}^\prime ).
\end{multline}
Rearranging:
\begin{multline}
    N_{\text{PDN},\text{PDN}} = i_{\text{PDN},\text{RMS}}i_{\text{PDN},\text{RMS}}^\prime \frac{\delta \nu}{(N+1) B_f} \\ \sum_{n = 0}^N \sum_{k = -B_f/\delta \nu}^{B_f/\delta \nu} \cos(2 \pi k \delta \nu t + \phi_{k,n} ) \sum_{m = 0}^N \sum_{j = -B_f/\delta \nu}^{B_f/\delta \nu} \cos(2 \pi j \delta \nu t + \phi_{j,m}^\prime ).
\end{multline}
Rearranging the summations:
\begin{multline}
    N_{\text{PDN},\text{PDN}} = i_{\text{PDN},\text{RMS}}i_{\text{PDN},\text{RMS}}^\prime \frac{\delta \nu}{(N+1) B_f} \\ \sum_{n = 0}^N \sum_{m = 0}^N \sum_{k = -B_f/\delta \nu}^{B_f/\delta \nu} \sum_{j = -B_f/\delta \nu}^{B_f/\delta \nu} \cos(2 \pi k \delta \nu t + \phi_{k,n} ) \cos(2 \pi j \delta \nu t + \phi_{j,m}^\prime ).
\end{multline}
Next, we split this photocurrent into sum and difference terms, accounting for the low-pass filter after the mixer with bandwidth $B_{f,m}$:
\begin{multline}
    N_{\text{PDN},\text{PDN}} = i_{\text{PDN},\text{RMS}}i_{\text{PDN},\text{RMS}}^\prime \frac{\delta \nu}{(N+1) B_f} \\ \bigg( \sum_{n = 0}^N \sum_{m = 0}^N \sum_{k = -B_f/\delta \nu}^{B_f/\delta \nu} \sum_{j = -B_f/\delta \nu}^{B_f/\delta \nu} \cos(2 \pi k \delta \nu t + \phi_{k,n} + 2 \pi j \delta \nu t + \phi_{j,m}^\prime ) \\ + \sum_{n = 0}^N \sum_{m = 0}^N \sum_{k = -B_f/\delta \nu}^{B_f/\delta \nu} \sum_{j = -B_f/\delta \nu, |j+k| < B_{f,m}/\delta \nu}^{B_f/\delta \nu} \cos(2 \pi k \delta \nu t + \phi_{k,n} -(2 \pi j \delta \nu t + \phi_{j,m}^\prime )) \bigg)
\end{multline}
Assuming that $B_{f,m} = B_f$, there are 3/4 the terms. We now take the root-mean-square and sum in quadrature.
\begin{multline}
    N_{\text{PDN},\text{PDN},\text{RMS}} = i_{\text{PDN},\text{RMS}}i_{\text{PDN},\text{RMS}}^\prime \frac{\delta \nu}{(N+1) B_f} \\ \sqrt{\bigg( \frac{3}{4} \sqrt{(N+1)^2 (2Bf/\delta \nu)^2} \frac{1}{\sqrt{2}}\bigg)^2 + \bigg(\frac{3}{4} \sqrt{(N+1)^2 (2Bf/\delta \nu)^2} \frac{1}{\sqrt{2}} \bigg)^2}
\end{multline}
Simplifying:
\begin{multline}
    N_{\text{PDN},\text{PDN},\text{RMS}} = i_{\text{PDN},\text{RMS}}i_{\text{PDN},\text{RMS}}^\prime \frac{\delta \nu}{(N+1) B_f} \\ \frac{1}{\sqrt{2}} \sqrt{ \bigg(\frac{3}{4}\bigg)^2 (N+1)^2 (2Bf/\delta \nu)^2 + \bigg(\frac{3}{4}\bigg)^2(N+1)^2 (2Bf/\delta \nu)^2}
\end{multline}
Further simplifying:
\begin{equation}
    N_{\text{PDN},\text{PDN},\text{RMS}} = i_{\text{PDN},\text{RMS}}i_{\text{PDN},\text{RMS}}^\prime \frac{\delta \nu}{(N+1) B_f} \frac{\sqrt{2}}{\sqrt{2}} \frac{3}{4} \sqrt{(N+1)^2 (2Bf/\delta \nu)^2}
\end{equation}
Evaluating the square root:
\begin{equation}
    N_{\text{PDN},\text{PDN},\text{RMS}} = i_{\text{PDN},\text{RMS}}i_{\text{PDN},\text{RMS}}^\prime \frac{\delta \nu}{(N+1) B_f} \frac{3}{4} (N+1) (2Bf/\delta \nu)
\end{equation}
Further simplifying:
\begin{equation}
    N_{\text{PDN},\text{PDN},\text{RMS}} = 2 \times \frac{3}{4} i_{\text{PDN},\text{RMS}}i_{\text{PDN},\text{RMS}}^\prime
\end{equation}
\paragraph{Total Noise}
Next, we consider all the noise together.  The total noise is:
\begin{equation}
N_\text{tot} = \sqrt{N_{\text{het},\text{RMS}}^2 + (N_{\text{PDN},\text{het},\text{RMS}})^2 + (N^\prime_{\text{PDN},het,\text{RMS}})^2 + N_{\text{PDN},\text{PDN},\text{RMS}}^2}
\end{equation}
Plugging in these terms, we have:
\begin{multline}
    N_\text{tot}^2 = \bigg(8 \times \frac{3}{4} (\frac{\eta e}{h \nu})^2 S_{det} \sqrt{P_t P_t^\prime} \times
    (N+1) \times B_f\bigg)^2 + \bigg(2\times \frac{3}{4} \frac{\eta e}{h \nu} \sqrt{4 S_{det} P_t} i_{\text{PDN},\text{RMS}}^\prime \sqrt{(N+1) B_f}\bigg)^2 \\ + \bigg(2\times \frac{3}{4} \frac{\eta e}{h \nu} \sqrt{4 S_{det} P_t^\prime} i_{\text{PDN},\text{RMS}} \sqrt{(N+1) B_f}\bigg)^2 + \bigg(2 \times \frac{3}{4} i_{\text{PDN},\text{RMS}}i_{\text{PDN},\text{RMS}}^\prime \bigg)^2
\end{multline}
To simplify, we assume that the photodetected noise on both combs is the same. We also assume that the comb power is the same $P_t = P_t^\prime$.
\begin{multline}
    N_\text{tot}^2 = \bigg( 2 \times \frac{3}{4} \bigg)^2 \Bigg[\bigg( (\frac{\eta e}{h \nu})^2 4 S_{det} P_t \times
    (N+1) \times B_f\bigg)^2 + \bigg( \frac{\eta e}{h \nu} \sqrt{4 S_{det} P_t} i_{\text{PDN},\text{RMS}} \sqrt{(N+1) B_f}\bigg)^2 \\ + \bigg( \frac{\eta e}{h \nu} \sqrt{4 S_{det} P_t} i_{\text{PDN},\text{RMS}} \sqrt{(N+1) B_f}\bigg)^2 + \bigg(i_{\text{PDN},\text{RMS}}i_{\text{PDN},\text{RMS}} \bigg)^2\Bigg]
\end{multline}
We recognize that this is a squared sum:
\begin{equation}
    N_\text{tot}^2 = \bigg( 2 \times \frac{3}{4} \bigg)^2 \Bigg[ (\frac{\eta e}{h \nu})^2 4 S_{det} P_t \times
    (N+1) \times B_f + i_{\text{PDN},\text{RMS}}^2\Bigg]^2
\end{equation}
Expanding out the noise current term in terms of the noise power spectral density, we have:
\begin{equation}
    N_\text{tot} = \bigg( 2 \times \frac{3}{4} \bigg) \Bigg[ (\frac{\eta e}{h \nu})^2 4 S_{det} P_t \times
    (N+1) \times B_f + (N+1)B_f S_{IN }\Bigg]
\end{equation}
If this is at the shot noise limit, then the total noise is:
\begin{equation}
    N_{\text{tot},SNL} = \bigg( 2 \times \frac{3}{4} \bigg) \Bigg[ (\frac{\eta e}{h \nu})^2 4 S_{det} (N+1) P_t
  B_f + 2 \frac{\eta e^2}{h \nu} (N+1) P_t B_f \Bigg]
\end{equation}
\paragraph{Calculating SNR at the Shot Noise Limit}

Recall that the signal is:
\begin{equation}
    A_\text{het, peak} = 8 \times \frac{3}{4}(\frac{\eta e}{h \nu})^2 S_{det} \sqrt{P_t P_t^\prime} \times (N+1) \times B_f
\end{equation}
We'll now make a change of variables of $S_{tot} = 2 S_{det}$, referring now to the total thermal PSD being measured over both balanced detectors. We'll also assume the same comb power, i.e., $P_t = P_t^\prime$. Thus:
\begin{equation}
    A_\text{het, peak} =  \frac{3}{4}(\frac{\eta e}{h \nu})^2 4S_{tot} P_t \times (N+1) \times B_f.
\end{equation}
The SNL noise can be written as:
\begin{equation}
    N_{\text{tot},SNL} = \frac{3}{4} \Bigg[ (\frac{\eta e}{h \nu})^2 4 S_{tot} (N+1) P_t
  B_f + 4 \frac{\eta e^2}{h \nu} (N+1) P_t B_f \Bigg].
\end{equation}
The SNL SNR is:
\begin{equation}
\text{SNR}_{SNL} = \frac{A_\text{het, peak}}{N_{\text{tot},SNL}} = \frac{\frac{3}{4}(\frac{\eta e}{h \nu})^2 4S_{tot} P_t \times (N+1) \times B_f}{\frac{3}{4} \Bigg[ (\frac{\eta e}{h \nu})^2 4 S_{tot} (N+1) P_t
  B_f + 4 \frac{\eta e^2}{h \nu} (N+1) P_t B_f \Bigg]}.
\end{equation}
Simplifying,
\begin{equation}
    \text{SNR}_{SNL} = \frac{\frac{\eta}{h \nu} S_{tot}}{\Bigg[ \frac{\eta}{h \nu} S_{tot} +  1 \Bigg]} = \frac{\eta \langle n \rangle}{\eta \langle n \rangle + 1}.
\end{equation}
Note that one could also go through this process with RIN following the formalism in the last section. In the main text we simply parameterize the RIN as part of the total noise, which is a multiple of the shot noise, hence the equation:
\begin{equation}
    \text{SNR}_\text{Tech} = \frac{\eta \langle n \rangle}{\eta \langle n \rangle + \chi}.
\end{equation}
Where $\chi$ is the multiple of the total noise including technical noise (such as RIN) over just the shot noise.

\section{Comparison With Other Spectroscopy Techniques}\label{supp: comparison}

\begin{figure*}[!ht]%
\centering
\includegraphics[width=1\textwidth]{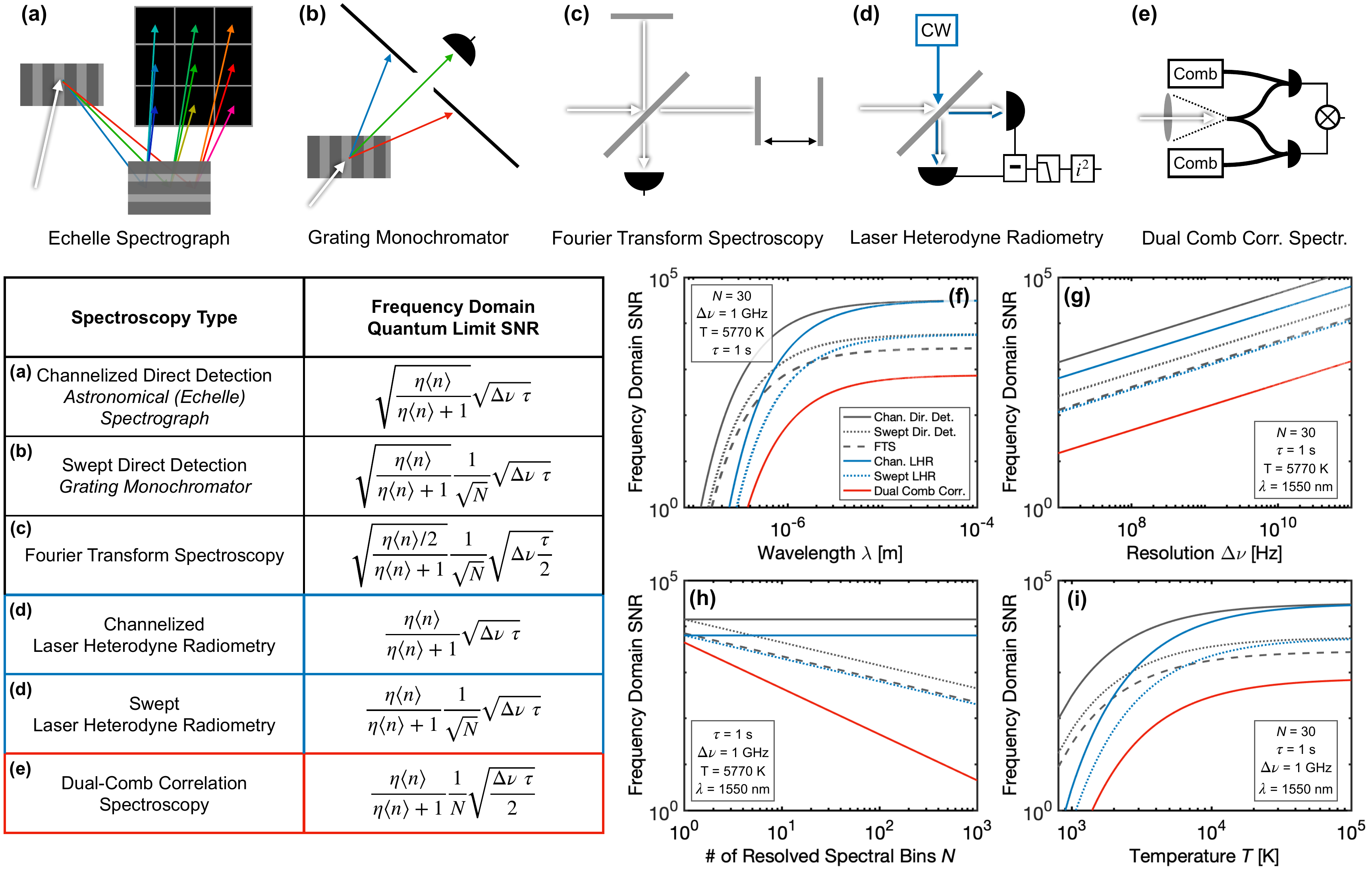}
\caption{Comparing SNR of thermal spectroscopy techniques \textbf{(a)} Echelle spectrograph diagram. \textbf{(b)} Grating monochromator diagram. \textbf{(c)} Fourier transform spectrometer diagram. \textbf{(d)} Laser heterodyne radiometry diagram. Channelized LHR would host many CW LOs and detectors. Swept LHR would sweep the frequency of the CW LO using one detector. \textbf{(e)} Dual comb correlation diagram. \textbf{(f)} Frequency domain SNR vs. wavelength \textbf{(g)} Frequency domain SNR vs. resolution \textbf{(h)} Frequency domain SNR vs. $N$, the number of resolved spectral bins \textbf{(i)} Frequency domain SNR vs black-body temperature. Note, $\langle n \rangle = (\exp{[h \nu / kT]}-1)^{-1}$. }\label{fig:comparison}
\end{figure*}

To evaluate the utility of DCCS, we compare the frequency domain SNR of DCCS (see Supplementary Note \ref{app: timetofreqSNR}) alongside existing methods for thermal light spectroscopy. These methods include channelized direct detection (echelle spectrographs--Fig. \ref{fig:comparison}a), where all resolved frequency bins are channelized onto as many photodetectors \cite{Zmuidzinas2003}; swept direct detection (grating monochromators--Fig. \ref{fig:comparison}b), where thermal light is separated on a grating and scanned across a single detector; shot noise-limited Fourier transform spectroscopy (FTS--Fig. \ref{fig:comparison}c) \cite{Boudreau2012}; channelized laser heterodyne radiometry (Fig. \ref{fig:comparison}d), where all resolved frequency bins are channelized onto as many photodetectors \cite{Zmuidzinas2003}; swept laser heterodyne radiometry (Fig. \ref{fig:comparison}d), where laser local oscillator light is scanned in frequency using a single detector; and DCCS (Fig. \ref{fig:comparison}e). Note that we consider the detection of a single mode of thermal light and thus do not treat the multiplex advantage possible in direct detection techniques. However, the most precise spectral measurements in direct detection also image a single mode due to dispersion-related uncertainty \cite{Crass2019}. Thus we believe that a single spatial mode-based comparison is fair and illustrative.

The SNR of all methods share the well-known square root scaling with averaging time and frequency resolution, but differ in scaling with respect to photon occupation $\langle n \rangle$ and number of resolved frequency bins $N$ (Fig. \ref{fig:comparison} Table). For swept techniques, FTS, and DCCS, $N$ is the number of resolved frequency bins per detector. Channelized techniques, however, host one resolved frequency bin per detector and thus do not scale with $N$.

We graphically compare the SNR as a function of wavelength ($\lambda$), resolution ($\Delta \nu$), number of resolved spectral bins ($N$), and black body temperature (T) in Fig. \ref{fig:comparison}.  In each of the separate plots, we scan one variable from standard conditions of $N = 30$, $\Delta \nu = 1 \text{ GHz}$, $\text{T} = 5770 \text{ K}$, $\tau = 1 \text{ s}$, $\lambda$ = 1550 nm. Note the total bandwidth measured is $N \Delta \nu = 30 \text{ GHz}$. In Fig. \ref{fig:comparison}f, we see that all techniques suffer SNR degradation at shorter wavelengths, but direct detection techniques fare better in the regime where $\langle n \rangle<1$ . This is due to direct detection techniques scaling as $\sqrt{\langle n \rangle}$ vs heterodyne techniques scaling as $\langle n \rangle$. At longer wavelengths, all techniques exhibit SNR of greater than $10^2$, sufficient for many applications. Square root scaling with spectral resolution $\Delta \nu$ is shown in Fig. \ref{fig:comparison}g. In Fig. \ref{fig:comparison}h we plot the SNR scaling with number of resolved spectral bins $N$, showing that all techniques are roughly equivalent when $N=1$, but DCCS shows a disadvantage at high $N$ due to $1/N$ scaling. One can think of this scaling as a temporal mode-mismatch penalty between the two comb LOs and the thermal light, or as an optical to RF spectral compression penalty. Fig. \ref{fig:comparison}i shows scaling with temperature, illustrating that cooler objects are better measured with direct detection methods due to $\sqrt{\langle n \rangle}$ scaling.

\section{Conversion of Time Domain SNR to Frequency Domain SNR}\label{app: timetofreqSNR}

The preceding SNR describes the single-shot time-domain SNR. To reach the SNR in the frequency domain over some time $\tau$, one must multiply this figure by the corresponding number of interferograms and then divide by the square root of the number of resolved frequency bins. This procedure can be found in the noise analysis of \cite{Newbury2010} as well as in \cite{Boudreau2012}, and it accounts for the difference in time vs frequency SNR. In DCCS, since the SNR in the time domain is always limited to 1 for thermal light, additional resolved frequency bins ``dilute'' the SNR in the frequency domain.

\section{Analysis of Resolvable Line Shift and Doppler Velocity Uncertainty}\label{app: lineshift}

The uncertainty in the position of a spectral absorption line ($\sigma_\nu$) is related to the uncertainty in the amplitude by the following relationship:
\begin{equation}
\sigma_\nu = \sigma_f/|df/d\nu|,
\end{equation}
where $\nu$ is the optical frequency and $f(\nu)$ is the spectral line function. One can rewrite this in terms of SNR and the line depth ($D$) of the absorption line as:
\begin{equation}
\sigma_\nu = \frac{1}{\text{SNR} \times D \times \frac{ds}{d\nu}},
\end{equation}
where $s(\nu)$ is the normalized spectral line function. Consider the Gaussian function $s(\nu) = e^{-\frac{1}{2}(\frac{\nu}{\mathcal{D}\nu})^2}$ , where $\mathcal{D}\nu$ is the line width. The derivative is:
\begin{equation}
\frac{ds}{d\nu} = \frac{-\nu}{\mathcal{D}\nu^2} e^{-\frac{1}{2}(\frac{\nu}{\mathcal{D}\nu})^2},
\end{equation}
and the uncertainty is then:
\begin{equation}
\sigma_\nu = \frac{1}{\text{SNR} \times D \times \frac{-\nu}{\mathcal{D}\nu^2} e^{-\frac{1}{2}(\frac{\nu}{\mathcal{D}\nu})^2}}.
\end{equation}
We are interested in the total uncertainty over a measurement resolving the whole line. The uncertainty of multiple measurements is:
\begin{equation}
\sigma_{tot} = \frac{1}{\sqrt{\sum_{i} \sigma_{\nu,i}^{-2}}} = \frac{1}{\sqrt{\sum_{i} (\text{SNR} \times D \times \frac{-\nu_i}{\mathcal{D}\nu^2} e^{-\frac{1}{2}(\frac{\nu_i}{\mathcal{D}\nu})^2})^2}},
\end{equation}
Simplifying,
\begin{equation}
\sigma_{tot} = \frac{1}{\frac{\text{SNR} \times D}{\mathcal{D}\nu^2}\sqrt{\sum_{i} (-\nu_i e^{-\frac{1}{2}(\frac{\nu_i}{\mathcal{D}\nu})^2})^2}} = \frac{1}{\frac{\text{SNR} \times D}{\mathcal{D}\nu^2}\sqrt{\sum_{n=1}^{N\rangle 3\mathcal{D}\nu/\Delta \nu} (-2n\Delta \nu e^{-\frac{1}{2}(\frac{n\Delta \nu}{\mathcal{D}\nu})^2})^2}},
\end{equation}
where one is assuming the capture of the line at greater than 3 standard deviations. The Doppler shift that can be resolved is $\delta V = \sigma_{tot}\times c / \nu_0$, where $c$ is the speed of light and $\nu_0$ is the center optical frequency.

In the main text we consider an efficiency of $\eta = 0.5$, a line depth $D$ of 0.5, a line width $\mathcal{D}\nu$ of 3 GHz, a spectral resolution $\Delta \nu$ of 1 GHz,  and a center optical frequency $\nu_0$ of 200 THz (near 1550 nm). A $8.9$ cm/s uncertainty occurs at approximately SNR = 22,000, which occurs at around 2 hours and 20 minutes.

A similar analysis is carried out in \cite{Bouchy2001}. Note that if one only cares about the line center, there is no benefit to over-resolving the line with higher resolution. However, other phenomena may be of interest, such as complex magnetohydrodynamics that reshape the line itself. In such cases, over resolving the line may provide crucial information that disentangles line center shifts from other phenomena.

\section{Sensitivity of DCCS on $\alpha$ Centauri A}\label{appx: sensitivityalphacentauri}

Although heterodyne-based detection of Sunlight requires only a millimeter-sized aperture to capture the full Planck-Law-limited mode of the Sun, current telescopes are not large enough to fully capture the mode for other Sun-like stars in the galaxy. An important question is what sensitivity we can expect for these other stars with DCCS.

Our nearest sun-like star, $\alpha$ Centauri A, provides an instructive case. The de-rating factor, or filling factor, is $\eta_{f} = A_d \Omega/\lambda^2$, where $A_d$ is the telescope aperture area, $\Omega$ is the solid angle subtended by the distant body, and $\lambda$ is the wavelength \cite{Blaney1975}. Note that the shot-noise-limited single-shot SNR is then $\eta_f \eta \langle n \rangle/(\eta_f \eta \langle n \rangle + 1)$. The solid angle subtended by $\alpha$ Centauri A is $2.7\times10^{-15}$ rad$^2$. For a 10 meter telescope operating at 1.5 µm, $\eta_{f} \approx 0.093$.

Given the parameters listed in Appendix \ref{app: lineshift} with this additional filling factor penalty, the required time to reach the precision to detect an exo-Earth is approximately $(1/0.093)^2 = 115$ times longer than when the star is fully resolved. This translates to a required measurement duration of nine days.

\section{Fringe Capture Speed for $\alpha$ Centauri A}\label{appx: fringealphacentauri}

In the example of $\alpha$ Centauri A, a telescope diameter of approximately 33 meters would be required to fully capture the mode, that is, to reach $\eta_f = 1$, and such telescopes are in development. However, as mentioned in Appendix \ref{appx: synthesis imaging}, one attractive alternative is to use an array of smaller and less costly telescopes to fill such an area. Atmospheric turbulence and changing path lengths over the atmospheric column pose challenges to such coherent synthesis, but if interferograms and fringes can be captured quickly enough, these path-length changes can be corrected for in the synthesis process.

Consider a 10 millisecond interval during which a fringe must be captured and thus a time-domain SNR greater than 1 must be realized. Again, consider the use of 10-meter telescopes. The single shot filling factor-derated SNR with $\eta_f = 0.093$ and further measurement inefficiency $\eta = 0.5$ is 0.011 (at 1550 nm for a 5700 K black body). At this single-shot SNR, 33,000 interferograms are required to reach an SNR of 2 sufficient for fringe capture and realignment. The main constraint for this acquisition speed is the resolution and optical bandwidth captured on a single detector--faster acquisition generally means that one must either coarsen the resolution or reduce the optical bandwidth. In this case, to reach 33,000 interferograms in 10 ms, a $\Delta f_r = 3.3$ MHz is required. At a repetition rate/spectral resolution of $f_r = 5$ GHz, the optical bandwidth that can be captured (while providing a unique one-to-one mapping between optical and RF \cite{Coddington2016}) on a single photodetector is still a relatively substantial $\Delta \nu_0 = f_r^2/2\Delta f_r = 3.8$ THz. As noted in the main text, it is generally preferable to allocate significantly less than a THz of optical bandwidth per detector to enhance the spectral signal-to-noise ratio (SNR). As a point of reference, telecommunications dense wavelength division multiplexing (DWDM) channels typically operate around 100 GHz.

It is also instructive to understand what size telescope would be needed given the use of 100 GHz DWDM channels. At 5 GHz resolution, this allows for a $\Delta f_r = 125$ MHz, or the capturing of 1,250,000 interferograms in 10 ms. Indeed, an even more conservative SNR of 3 can be reached in 10 ms with a measurement inefficiency of $\eta = 0.5$ and a filling factor-related inefficiency of $\eta_f = 0.023$, equivalent to a 5-meter telescope.

It is important to note that direct detection-based phase correction can also be used to further increase the coherent averaging time, thus allowing increased sensitivity beyond that discussed above. In conjunction with a heterodyne array, such a direct detection-based phase correction system need only connect nearest neighbors, avoiding both over-resolution issues as well as sensitivity loss due to light splitting, and could be fiber-integrated \cite{Ireland2014}.

\end{appendices}


\bibliography{main.bib}


\end{document}